\documentclass[ 12pt, letterpaper]{article}%
\usepackage{fullpage}
\usepackage{amsmath}
\usepackage{graphicx}%
\usepackage{amsfonts}%
\usepackage{amssymb}
\usepackage{fancyhdr}
\usepackage{isomath}
\usepackage{amsbsy}
\usepackage{amscd}
\usepackage{verbatim}
\usepackage{euscript}
\usepackage{alltt}
\usepackage{stmaryrd}
\usepackage{float}
\usepackage{appendix}
\usepackage{caption}
\usepackage[english]{babel}

\usepackage{subfigure}
\usepackage{tikz}
\usetikzlibrary{shapes,arrows}

\usepackage{amsmath}
\usepackage{amsbsy}
\usepackage{amssymb}
\usepackage{amscd}
\usepackage{amsfonts}

\newcommand{\bfa}{{\mathbold a}}
\newcommand{\bfb}{{\mathbold b}}

\newcommand{\bfd}{{\mathbold d}}
\newcommand{\bfe}{{\mathbold e}}
\newcommand{\bff}{{\mathbold f}}

\newcommand{\bfi}{{\mathbold i}}

\newcommand{\bfm}{{\mathbold m}}
\newcommand{\bfn}{{\mathbold n}}

\newcommand{\bfp}{{\mathbold p}}

\newcommand{\bfv}{{\mathbold v}}

\newcommand{\bfx}{{\mathbold x}}

\newcommand{\bfA}{{\mathbold A}}
\newcommand{\bfB}{{\mathbold B}}

\newcommand{\bfD}{{\mathbold D}}

\newcommand{\bfI}{{\mathbold I}}

\newcommand{\bfK}{{\mathbold K}}

\newcommand{\bfM}{{\mathbold M}}

\newcommand{\bfR}{{\mathbold R}}

\newcommand{\bfT}{{\mathbold T}}

\newcommand{\bfV}{{\mathbold V}}

\newcommand{\bfX}{{\mathbold X}}

\newcommand{\beq}{\begin{equation}}
\newcommand{\eeq}{\end{equation}}
\newcommand{\beqs}{\begin{eqnarray}}
\newcommand{\eeqs}{\end{eqnarray}}
\newcommand{\beql}{\begin{equation} \label}

\newcommand{\bfnu}{\mathbold{\nu}}

\newcommand{\bfomega}{\mathbold{\omega}}

\newcommand{\bfLambda}{\mathbold{\Lambda}}

\newcommand{\bfOmega}{\mathbold{\Omega}}

\newcommand{\grad}{\mathop{\rm grad}\nolimits}
\newcommand{\divergence}{\mathop{\rm div}\nolimits}
\newcommand{\curl}{\mathop{\rm curl}\nolimits}

\usepackage[margin=1in]{geometry}
\usepackage{titlesec}

\begin{document}

\title{Computational modeling of tactoid dynamics in chromonic liquid crystals}
\author{Chiqun Zhang$^1$, Amit Acharya$^1$, \\ Noel J. Walkington$^1$, Oleg D. Lavrentovich$^2$ \\
$^1$Carnegie Mellon University, Pittsburgh, USA\\
$^2$Kent State University, Kent, USA}

\maketitle


\begin{abstract}

\noindent Motivated by recent experiments, the isotropic-nematic phase transition in chromonic liquid crystals is studied. As temperature decreases, nematic nuclei nucleate, grow, and coalesce, giving rise to tactoid microstructures in an isotropic liquid. These tactoids produce topological defects at domain junctions (disclinations in the bulk or point defects on the surface). We simulate such tactoid equilibria and their coarsening dynamics with a model using degree of order, a variable length director, and an interfacial normal as state descriptors. We adopt Ericksen's work and introduce an augmented Oseen-Frank energy, with non-convexity in both interfacial energy and the dependence of the energy on the degree of order. A gradient flow dynamics of this energy does not succeed in reproducing some simple expected feature of tactoid dynamics. Therefore, a strategy is devised based on continuum kinematics and thermodynamics to represent such features. The model is used to predict tactoid nucleation, expansion, and coalescence during the process of phase transition. We reproduce observed behaviors in experiments and perform an experimentally testable parametric study of the effect of bulk elastic and tactoid interfacial energy parameters on the interaction of interfacial and bulk fields in the tactoids.
\end{abstract}

\section{Introduction}

Liquid crystals (LC) are a state of matter with long-range orientational order and complete (nematic) or partial (smectics, columnar phases) absence of long-range positional order of `building units' (molecules, viruses, aggregates, etc.). Liquid crystals can flow like viscous liquids, and also possess features that are characteristic of solid crystals, such as elasticity and birefringence. In the simplest liquid crystalline phase, called the nematic,  the molecules have no positional order but tend to point in the same direction. In this work, we focus on a nematic lyotropic liquid crystal (LCLC) that possesses a broad biphasic region of coexisting nematic and isotropic phases \cite{kim2013morphogenesis}.

LCLCs are formed by water-based dispersions of organic molecules, see the recent reviews \cite{lydon2011chromonic, collings2015nature, li2012liquid}. The molecules are of a rigid disc-like or plank-like shape with polar groups at the periphery. Once in water, they form elongated aggregates by stacking on top of each other. The aggregates elongate as the concentration is increased and the temperature is reduced, which allows one to trigger phase transitions in the system by changing either the temperature or concentration \cite{nastishin2004pretransitional, park2011condensation}. In particular, the temperature changes can trigger a first order isotropic-nematic (I-N) phase transition of the LCLC. As the temperature increases, the nematic liquid crystal loses orientational order and transits to the isotropic phase, with molecular aggregates being short and oriented randomly. On the contrary, if the temperature decreases, the isotropic phase transits to the nematic phase. Both phase transitions occur through nucleation of the so-called tactoids, representing inclusions of one phase in the other \cite{kim2013morphogenesis, tortora2010, nastishin2005, tortora2011chiral}. Tactoids of the nematic phase nucleating upon cooling are called positive tactoids \cite{nastishin2005} and are the subject of the present work. Tactoids of the isotropic phase nucleating in the nematic background upon heating are called negative tactoids \cite{nastishin2005}. If the temperature is fixed in the range in which the two phases coexist, these tactoids expand and merge. The uniaxial nematic phase allows three types of topologically stable defects: linear disclinations, point defects-hedgehogs and point defect-boojums; the latter can exist only at the surface of the nematic \cite{kleman2007soft, chuang1991cosmology, Bowick943, vachaspati1991formation}. In confined volumes, such as droplets and tactoids, some of the topological defects correspond to the equilibrium state of the system, thanks to the anisotropic surface tension that sets a well-defined angle between the director and the normal to the interface \cite{volovik1983topological}.

The principal objectives of this work are to:
\begin{itemize}
\item derive a practical equation of evolution for the degree of orientation based on kinematics and thermodynamics;
\item introduce a dynamic model for the nematic-isotropic phase transition of LCLC with an augmented Oseen-Frank energy and non-convex interfacial energy;
\item demonstrate the capability of the proposed dynamical model by analyzing the results of static equilibrium and the dynamic behaviors.
\end{itemize}

The main experimental observations and applications of LCLC and their computation are reviewed in \cite{kim2013morphogenesis, li2012liquid, lydon2011chromonic, collings2015nature}. Currently, there is an extensive database on the principal material parameters of the LCLCs and defects in them. All three bulk elastic constants (for splay $K_{11}$, twist $K_{22}$ and bend $K_{33}$) have been measured for two main representatives of LCLCs \cite{zhou2012elasticity, zhou2017elasticity, zhou2017ionic}. It was found that the elastic constants of bend and splay can be tuned in a broad range, from a few pN to 70 pN, by changing temperature or the chemical composition of the system (e.g., by adding salts \cite{zhou2017ionic}). The director of LCLCs can align either parallel to the interface with an adjacent medium \cite{schneider2005oriented} or in a perpendicular fashion, with possible transitions between these two states \cite{nazarenko2010surface}. At the interface with its own isotropic melt, the director of a nematic LCLC aligns parallel to it \cite{tortora2010self}. The interfacial surface tension at the isotropic-nematic interface was estimated to be on the order of $10^{-4}$ $J/m^2$ \cite{kim2013morphogenesis}. The defect cores of disclinations in LCLCs extend over long distances (microns and even tens of microns), much larger than the cores of disclinations in thermotropic liquid crystals \cite{zhou2017fine}.

\begin{figure}[H]
\centering
\includegraphics[width=0.8\textwidth]{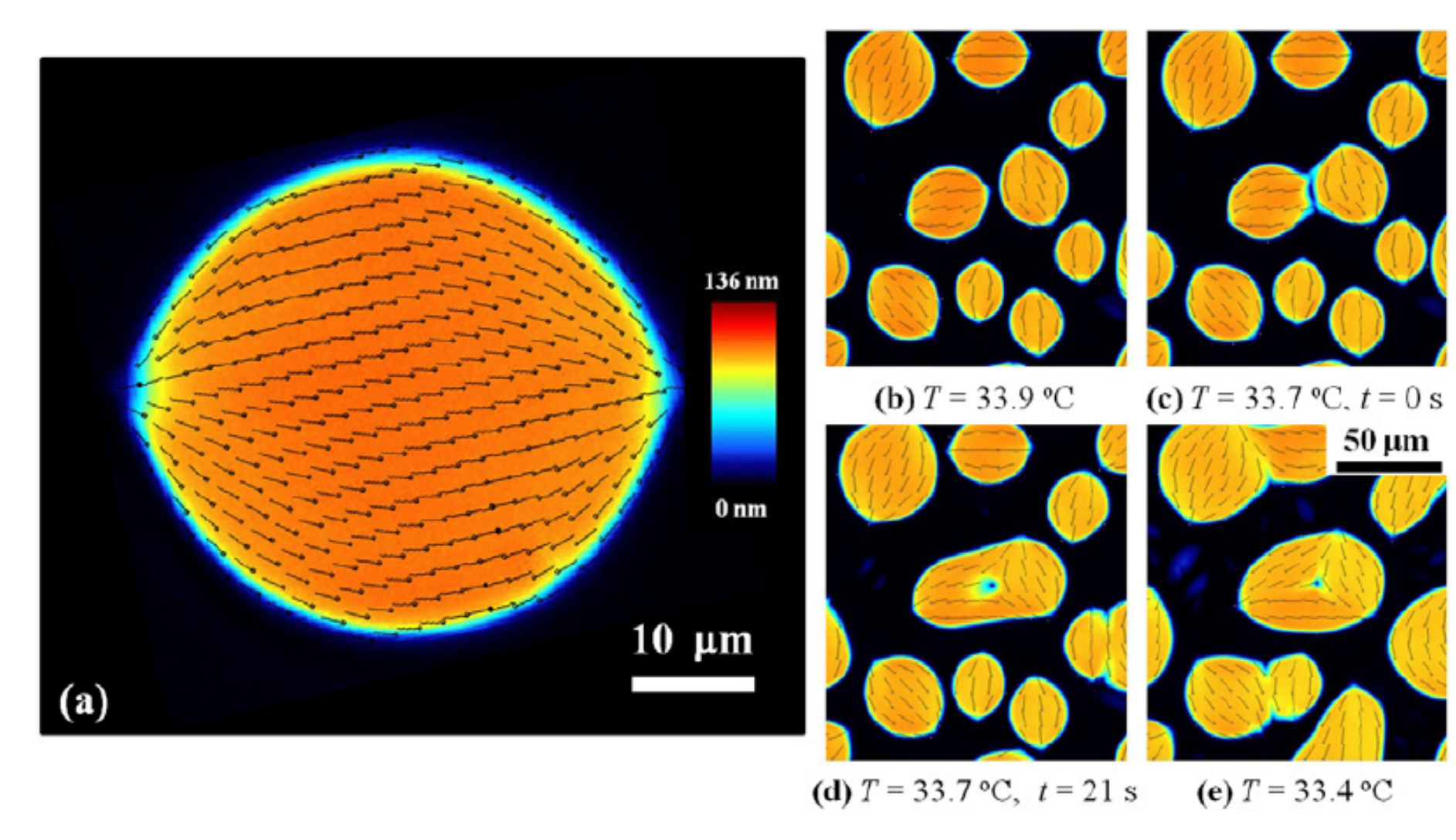}
\caption{Experiment observations of isotropic-nematic phase transition from \cite{kim2013morphogenesis}.}
\label{fig:intro_experiment}
\end{figure}

In this work, we are primary interested in the observations reported in \cite{kim2013morphogenesis} to develop a model for understanding the behavior of tactoids during the isotropic-nematic transformation. The isotropic-nematic interface in LCLC favors the director to be tangential to the tactoid interface. Fig. \ref{fig:intro_experiment} shows the experimental observations of the isotropic-nematic phase transition from \cite{kim2013morphogenesis}.  Fig. \ref{fig:intro_experiment}(a) shows a single tactoid, where the black color represents the isotropic phase while the orange color represents the nematic phase. The black arrows inside the tactoid represent the director field. Nontrivial morphologies of tactoids with surface cusps and director fields are observed. Due to the surface anisotropy, cusps are associated with surface defects called boojums, as shown in Fig. \ref{fig:intro_experiment}(a).  Fig. \ref{fig:intro_experiment}(b) to \ref{fig:intro_experiment}(e) represent the phase transition process from the isotropic to the nematic phase, where the nematic tactoids expand and merge.  Merging tactoids often produce disclinations via the Kibble mechanism \cite{chuang1991cosmology, Bowick943, kibble1976}, as shown in Fig. \ref{fig:intro_experiment}(e), where a strength $-\frac{1}{2}$ disclination is formed at the point where tactoids merge. In addition, integer strength disclinations are stable only when their cores constitute a large isotropic inclusion; otherwise, as demonstrated experimentally and analytically by Y.-K. Kim et al \cite{kim2013morphogenesis} and numerically in \cite{zhang2016non}, the integer strength disclinations split into pairs of half-integer ones. The motion of an interface between a nematic liquid crystal phase and the isotropic phase is investigated with a Ginzburg-Landau equation in \cite{popa1996kinetics}. The confinement of the director field for a spherical particle that explains the observation of a Saturn ring is studied in \cite{grollau2003spherical}.

In studies of nematic liquid crystals, a classical convention is to represent the local orientational order by a unit-length director field \cite{virga1995variational, ericksen1991liquid}. Oseen and Frank developed an energy density of nematic liquid crystals, with constants representing different director deformations \cite{frank1958liquid, oseen1933theory}. The existence and partial regularity theory of some boundary-value problems based on Oseen-Frank energy density are discussed in \cite{hardt1986existence}.  The Oseen-Frank energy can be augmented by adding an additional surface energy density to represent the interaction between the LC and an adjacent medium; a common form of such a surface energy density is the Rapini-Papoular surface energy.

In this paper we develop a computational model for the isotropic-nematic phase transition accounting for interfacial energy as an enhancement of Ericksen's variable degree of order $(s,\bfn)$ model \cite{ericksen1991liquid}. We introduce the pair $(s,\bfd)$ with $(\bfd = s\bfn)$. The state variable $s$ has the meaning of the degree of order parameter in Ericksen's model \cite{ericksen1991liquid} and $\bfd$ serves for the director whose magnitude is constrained to be equal to $|s|$. Thus, the director is of unit length in the nematic phase, it vanishes in the isotropic phase, and it is of variable length at interfaces between the two phases. This practical device of replacing $\bfn$ by $\bfd$ is essential in terms of having a setting that is well-posed for computations of a time-dependent nonlinear theory, since leaving the value of the director field undefined in parts of the domain, that furthermore evolve in time, does not lead to unique evolution and simply cannot be practically implemented.

The rest of this paper is organized as follows: In Section \ref{sec:notation}, we outline our notation and terminology. In Section \ref{sec:model}, a dynamic model for the phase transition process based on kinematics as well as thermodynamics is derived. In Section \ref{sec:results}, the results of equilibrium and dynamic behaviors are shown and discussed. The significance of the dynamic model is demonstrated and explained. In Section \ref{sec:para}, we report on a preliminary parametric study of material constants in the model. We end with some concluding remarks in Section \ref{sec:conclusion}.

\section{Notation and terminology} \label{sec:notation}

The condition that $a$ is defined to be $b$ is indicated by the statement $a := b$. The Einstein summation convention is implied unless specified otherwise. The symbol $\bfA \bfb$ denotes the action of a tensor $\bfA$ on a vector $\bfb$, producing a vector. In the sequel, $\bfa\cdot\bfb$ represents the inner product of two vectors $\bfa$ and $\bfb$; the symbol $\bfA\bfB$ represents tensor multiplication of the second-order tensors $\bfA$ and $\bfB$.

The symbol $\divergence$ represents the divergence and $\grad$ represents the gradient. In this paper all tensor or vector indices are written with respect to the basis  $\bfe_i$, $\bfi$=1 to 3, of a rectangular Cartesian coordinate system. The following component-form notation holds: 
\begin{equation*}
 \begin{split}
    \left(\bfa\times\bfb\right)_{i} &= e_{ijk} a_{j} b_{k}    \\
    \left(\curl\bfa\right)_{i} &= e_{ijk} a_{k,j}             \\
    \left(\divergence\bfA\right)_{i} &= A_{ij,j}                \\
    \left( \bfA :\bfB \right) &= A_{ij} B_{ij}
 \end{split}
\end{equation*}
where $e_{mjk}$ is a component of the alternating tensor $\bfX$.

The following list describes some of the mathematical symbols we use in this work:

$\bfn$: the unit vector field representing the director

$s$: the degree of orientation, $s=0$ represents the isotropic phase while $s=1$ represents the nematic phase

$\bfd$: the alternative vector field representing the director with $\bfd=s\bfn$

$\psi$: the free energy density

\section{Derivation of dynamic model} \label{sec:model}

\subsection{$s$ evolution equation in Ericksen-Leslie model}

In \cite{ericksen1991liquid}, Ericksen introduced a variable degree of orientation $s$ to represent different phase states of a liquid crystal. In his model, $s=0$ represents the isotropic phase and $s=1$, the nematic phase. Also, a unit length vector field is introduced to represent the director field, denoted as $\bfn$. In Ericksen's model, the balance law to derive the $s$ evolution equation is given as 
\begin{equation*}
\dot{P}  = div (\bfT) + G^I + G^E,
\end{equation*}
where $\psi$ is free energy density, $P$ is a generalized momentum with $P = \partial \psi / \partial s$, $\bfT$ is a generalized stress, $G^I$ represents a kind of internal body force with $G^I = -\partial \psi / \partial s + \hat{G}$, and $G^E$ is an external effect. Assuming the free energy density $\psi$ depends on $(s, grad s, \bfn, grad \bfn)$, and following the argument in \cite{ericksen1991liquid}, we have
\begin{eqnarray*}
&\dot{\overline{\frac{\partial \psi}{\partial s}}} = div (\frac{\partial \psi}{\partial \nabla s}) -  \frac{\partial \psi}{\partial s} + \hat{G} + G^E \\
&\Rightarrow \frac{\partial^2 \psi}{\partial s^2} \dot{s} + \frac{\partial^2 \psi}{\partial s \partial grad s}\cdot \dot{\overline{grad s}} + \frac{\partial^2 \psi}{\partial s \partial \bfn}\cdot \dot{\bfn} + \frac{\partial^2 \psi}{\partial s \partial grad \bfn}:\dot{\overline{grad \bfn}} = div (\frac{\partial \psi}{\partial grad s}) -  \frac{\partial \psi}{\partial s} + \hat{G} + G^E.
\end{eqnarray*}
After rearranging the terms, $s$ evolution equation in Ericksen's model can be written as  
\begin{eqnarray*}
&(\frac{\partial^2 \psi}{\partial s^2} ) \dot{s} + \frac{\partial^2 \psi}{\partial s \partial grad s}\cdot \dot{\overline{grad s}} = \\
&div (\frac{\partial \psi}{\partial grad s}) -  \frac{\partial \psi}{\partial s} + \hat{G} + G^E  - \frac{\partial^2 \psi}{\partial s \partial \bfn}\cdot \dot{\bfn} - \frac{\partial^2 \psi}{\partial s \partial grad \bfn}:\dot{\overline{grad \bfn}}.
\end{eqnarray*}
In this work, we would like to adopt a simpler evolution statement since the fundamental basis for Ericksen's balance law $\dot{P} = div (\bfT) + G^I + G^E$ is not clear to us. In particular, the coefficient $\frac{\partial^2 \psi}{\partial s^2}$ may change sign as the dependence on $s$ of the energy is nonconvex.

\subsection{Motivation and derivation of $s$ evolution} \label{sec:evols}

\begin{figure}[H]
\centering
\includegraphics[width=0.3\textwidth]{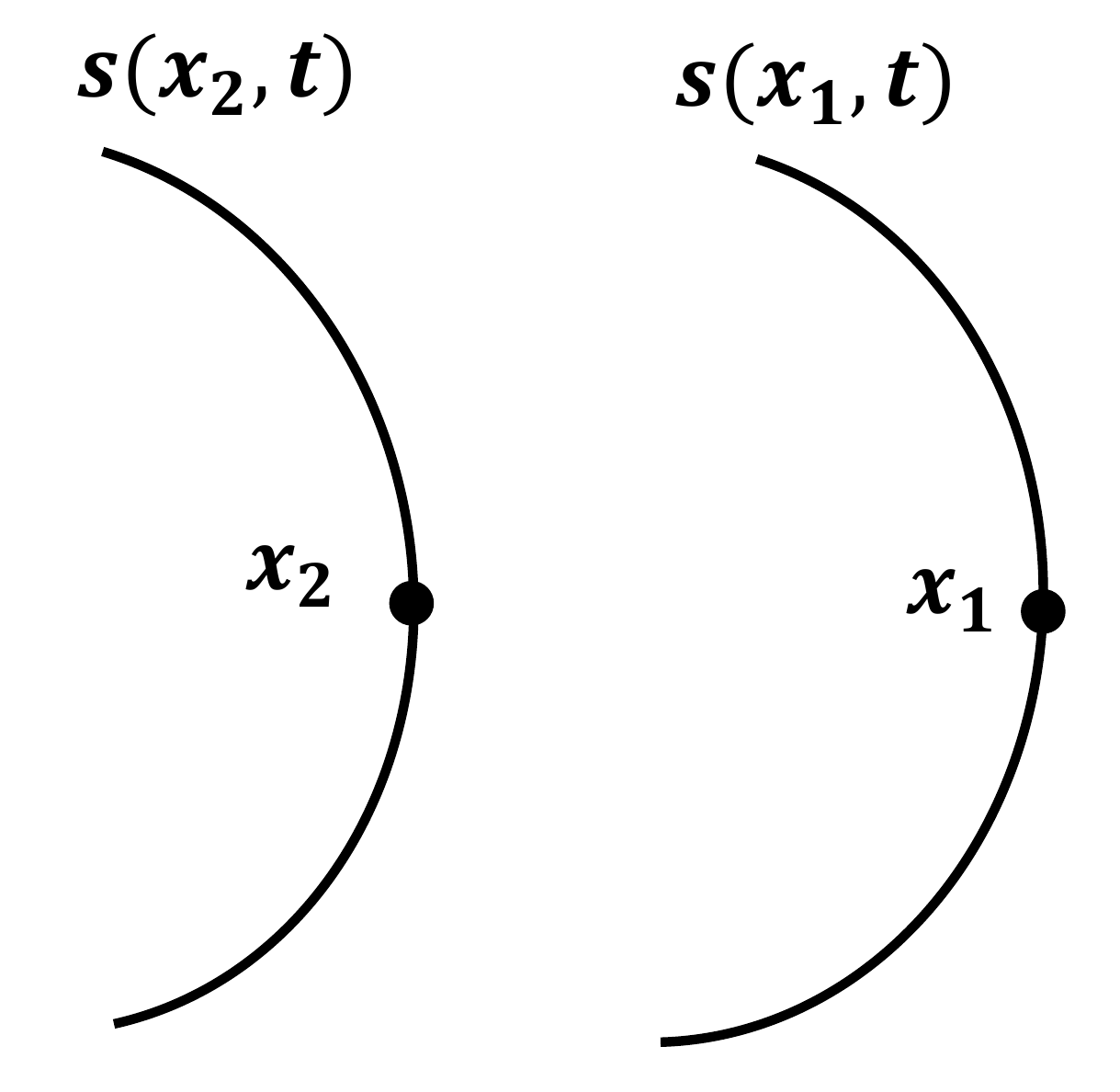}
\caption{A levelset of $s$ moving from $\bfx_2$ to $\bfx_1$ during $\Delta t$.}
\label{fig:evol_levelset}
\end{figure}

We derive a practical model for tactoid and isotropic-nematic phase transition dynamics based on continuum kinematics and thermodynamics. To get the evolution equation for $s$, suppose there is a level set of $s$ with normal velocity field $\bfV^{(s)}$ along it, traveling from $\bfx_2$ to $\bfx_1$ during a time interval $\Delta t$, as shown in Fig. \ref{fig:evol_levelset}. The time derivative of $s$ at $\bfx_1$ and $t$ is 
\begin{equation}
\frac{\partial s}{\partial t} = \lim_{\Delta t \to 0} \frac{s(\bfx_1, t+\Delta t)-s(\bfx_1, t)}{\Delta t}.
\end{equation}

Since the level set of s travels from $\bfx_2$ to $\bfx_1$ during the time interval $\Delta t$, $s(\bfx_1, t+\Delta t) = s(\bfx_2 (\Delta t), t)$. Thus, $\frac{\partial s}{\partial t}$ may also be expressed as   
\begin{equation}\label{eqn:dsdt}
\frac{\partial s}{\partial t} = \lim_{\Delta t \to 0} \frac{s(\bfx_2(\Delta t),t)-s(\bfx_1, t)}{\Delta t}.
\end{equation}

Assuming $s$ is differentiable in its arguments and writing the derivative in the first argument as $grad s$, we have
\begin{equation}\label{eqn:s_evol_dsdt}
\begin{split}
s(\bfx_2(\Delta t),t) - s(\bfx_1,t) = grad s (\bfx_1,t)[\bfx_2(\Delta t) -\bfx_1]+o(\bfx_2(\Delta t)-\bfx_1) \\
\Rightarrow \frac{s(\bfx_2(\Delta t),t) - s(\bfx_1,t)}{\Delta t} = grad s(\bfx_1,t)\frac{1}{\Delta t}[\bfx_2(\Delta t)-\bfx_1]+\frac{1}{\Delta t}o(\bfx_2(\Delta t)-\bfx_1).
\end{split}
\end{equation}

Substitute \eqref{eqn:s_evol_dsdt} in \eqref{eqn:dsdt}, we have 
\begin{equation}\label{eqn:dsdt2}
\frac{\partial s}{\partial t} = grad s (\bfx_1,t) \lim_{\Delta t \to 0} \frac{\bfx_2(\Delta t)-\bfx_1}{\Delta t}+ \lim_{\Delta t \to 0} \frac{o(\bfx_2(\Delta t)-\bfx_1)}{\Delta t}.
\end{equation}
Denote $\bfV^{(s)}$ as the velocity of movement of a level set of $s$, $\bfV^{(s)} = \lim_{\Delta t \to 0} \frac{\bfx_1-\bfx_2(\Delta t)}{\Delta t}$. Since
\[
\lim_{\Delta t \to 0} \left| \frac{o(\bfx_2(\Delta t) -\bfx_1)}{\Delta t} \right| = \lim_{\Delta t \to 0} \frac{|o(\bfx_2(\Delta t)-\bfx_1)|}{|\bfx_2(\Delta t)-\bfx_1|}\frac{|\bfx_2(\Delta t)-\bfx_1|}{\Delta t} = -0\cdot |\bfV^{(s)} |= 0.
\]
\eqref{eqn:dsdt2} becomes 
\[
\frac{\partial s}{\partial t} = -grad s \cdot \bfV^{(s)}.
\]

If the material velocity is $\bfv$ and the change in the value of $s$ at $\bfx_1$ arises from factors more than the pure advection of the value of $s$ from $\bfx_2$ to $\bfx_1$ due to material motion, then we assign the rest of this change as occurring due to the progress of the phase transition front. In general, we can decompose $\bfV^{(s)} = \bfv + \bfV$, where $\bfV$ is the phase front velocity relative to the material and $\bfv$ is the material velocity. Recall that the material time derivative of $s$ is defined as 
\begin{eqnarray*}
\dot{s} := \frac{ds}{dt} = \frac{\partial s}{\partial t} + grad s \cdot \bfv;
\end{eqnarray*}
therefore, the $s$ evolution is given as 
\begin{equation}\label{eqn:sdot}
 \dot{s}  = -grad s \cdot \bfV.
\end{equation}

In particular, there are two special cases:
\begin{itemize}
\item Suppose this velocity was purely due to $s$ being transported by the material velocity $\bfv$. Then we have that $\dot{s} = \frac{\partial s}{\partial t} + \frac{\partial s}{\partial \bfx}\cdot \bfv = 0$. 
\item If there is no material velocity but transport is only due to motion of the phase front, then $\bfV^{(s)}$ is just the speed of the phase front transition $\bfV$.
\end{itemize}

To get an explicit form of the phase front velocity $\bfV$, assume the free energy density per unit mass takes the form $\psi(\bfn, grad \bfn, s, grad s)$. Following \cite{leslie1992continuum, acharya2013continuum}, take the external power as
\begin{equation*}
P(t) =\int_{\partial V} (\bfLambda\bfnu)\cdot \bfomega da + \int_V \rho \bfK \cdot \bfomega dv,
\end{equation*} 
where $\bfLambda$ is the couple stress tensor, $\bfK$ is the external body moment per unit mass, $\bfnu$ is the unit normal vector on the boundary of the body, and $\bfomega$ is the director angular velocity (we have ignored material motion for simplicity). Applying the divergence theorem,we have 
\begin{equation*}
\int_{\partial V}(\bfLambda \bfnu) \cdot \bfomega da = \int_{\partial V} \Lambda_{ij}\omega_i \nu_j da = \int_V (\Lambda_{ij,j}\omega_i + \Lambda_{ij}\omega_{i,j})dv.
\end{equation*} 
Thus, the external power $P$ can be written as
\begin{equation*}
P(t) = \int_V [div \bfLambda + \rho \bfK]\cdot \bfomega dv + \int_V \bfLambda : \bfM dv,
\end{equation*} 
where $\bfM$ is defined as director angular velocity gradient $\bfM = grad \bfomega$. Recall that the balance law of angular momentum reads as
\begin{equation*}
div \bfLambda + \rho \bfK = \bf0,
\end{equation*}
leading to 
\begin{eqnarray*}
P(t) = \int_V \bfLambda:\bfM dv.
\end{eqnarray*}

In addition, the second law of thermodynamics requires the dissipation to be equal or larger than zero, which is given as 
\begin{eqnarray} 
&\int_V[\bfLambda :\bfM] - \rho \dot{\psi}] dv \ge 0  \label{eqn:s_dissipation} \\
&\Rightarrow \int_V \left[ \Lambda_{ij}\omega_{i,j} -\rho \frac{\partial \psi}{\partial n_i} \dot{n_i} - \rho \frac{\partial \psi}{\partial (n_{i,j})} \dot{\overline{n_{i,j}}} - \rho \frac{\partial \psi}{\partial s}\dot{s} - \rho \frac{\partial \psi}{\partial (s_{,j})}\dot{\overline{s_{,j}}}\right]dv \ge 0 \nonumber.
\end{eqnarray}
As flow is ignored for the moment, the inequality takes the form
\begin{eqnarray*}
\int_V \left[\Lambda_{ij} \omega_{i,j} - \rho \frac{\partial \psi}{\partial n_i} \dot{n_i}- \rho \frac{\partial \psi}{\partial (n_{i,j})}\dot{n}_{i,j} - \rho \frac{\partial \psi}{\partial s} \dot{s} - \rho \frac{\partial  \psi}{\partial (s_{,j})}\dot{s}_{,j}\right] dv \ge 0 \\
\Rightarrow \int_V \left[\Lambda_{ij} \omega_{i,j} - \rho \frac{\partial \psi}{\partial n_i} (\omega \times n)_i -\rho \frac{\partial \psi}{\partial (n_{i,j})} (\omega \times n)_{i,j} - \rho \frac{\partial \psi}{\partial s} \dot{s} + \rho (\frac{\partial \psi}{\partial (s_{,j})})_{,j} \dot{s} \right] dv  \\-
\int_{\partial V} \rho \frac{\partial \psi}{\partial (s_{,j})} \dot{s} \nu_j da \ge 0.
\end{eqnarray*}
Defining the couple stress $\bfLambda$ as
\begin{equation*}
\Lambda_{ij} := \rho e_{inm}n_n \frac{\partial \psi}{\partial n_{m,j}},
\end{equation*}
and applying the Ericksen identity \cite{ericksen1961conservation} as
\begin{equation*}
\left(\frac{\partial \psi}{\partial \bfn} \otimes \bfn + \frac{\partial \psi}{\partial grad \bfn}(grad \bfn)^\intercal + \left(\frac{\partial \psi}{\partial grad \bfn}\right)^\intercal grad\bfn\right)_{skew} = \bf0,
\end{equation*}
we obtain 
\begin{equation*}
\Lambda_{ij} \omega_{i,j} - \rho \frac{\partial \psi}{\partial n_i} (\omega \times n)_i - \rho \frac{\partial \psi}{\partial (n_{i,j})} (\omega \times n)_{i,j} = 0.
\end{equation*}
Then the dissipation inequality becomes
\begin{equation*} \label{eqn:dissipation}
\int_V\left[- \rho \frac{\partial \psi}{\partial s} \dot{s} + \rho \left(\frac{\partial \psi}{\partial (s_{,j})}\right)_{,j} \dot{s} \right] dv - \int_{\partial V} \rho \frac{\partial \psi}{\partial (s_{,j})} \dot{s} \nu_j da \ge 0.
\end{equation*}
To fulfill this inequality, recalling (\ref{eqn:sdot}) that $\dot{s} =  -grad s\cdot \bfV$, one requires
\begin{eqnarray*}
-\left[\rho \frac{\partial \psi}{\partial s} - \rho \left(\frac{\partial \psi}{\partial (s_{,j})}\right)_{,j}\right]s_{,i}V_i &\ge& 0 \quad \text{at interior points} \\
-\rho \frac{\partial \psi}{\partial (s_{,j})}\nu_j s_{,i}V_i &\ge& 0 \quad \text{at points on boundary}.
\end{eqnarray*}
Therefore, the choice of $\bfV^B$ on the boundary pointing in the direction of 
\begin{equation*}
-\rho \left(\frac{\partial \psi}{\partial (grad s)}\cdot \bfnu \right) grad s,
\end{equation*}
and $\bfV^I$ in the interior pointing in the direction of 
\begin{equation*}
-\left[\rho \frac{\partial \psi}{\partial s} - \rho div \left(\frac{\partial \psi}{\partial (grad s)}\right)\right]grad s
\end{equation*}
satisfy the non-negative dissipative requirement. In particular, $\bfV^I$ in the interior may be further assumed as 
\begin{equation*}
\bfV^I = -\frac{grad s}{B_m |grad s|^m}\left[-\rho div\left(\frac{\partial \psi}{\partial (grad s)}\right) + \rho \frac{\partial \psi}{\partial s}\right].
\end{equation*}
where $B_m$ is a material constant required on dimensional grounds related to `drag', and $m$ is  a parameter representing different scenarios, which can be $0$, $1$ and $2$. With $\dot{s} = -grad s \cdot \bfV$, the evolution equation of $s$ can be written as 
\begin{equation} \label{eqn:s}
\dot{s}  = \frac{1}{B_m} |grad s|^{2-m} \rho \left[-\frac{\partial \psi}{\partial s} + div \left(\frac{\partial \psi}{\partial (grad s)}\right)\right].
\end{equation}
$m=0$ is the \emph{simplest} natural choice representing a linear kinetic assumption. $m=2$ corresponds to the evolution equation derived from the gradient flow method. To this is appended the balance laws of linear momentum and angular momentum, utilizing the constitutive equations for couple stress and stress, the latter arising from the thermodynamic procedure above when flow is included \cite{stewart2004static}.

Another way to obtain the $s$ evolution equation is the gradient flow method. The gradient flow dynamics (for a non-conserved quantity) assumes that all information on evolution is directly available (up to a material parameter) once the energy function is known. Consider the total energy 
\begin{equation*}
E = \int_V \rho\psi(\bfn, grad \bfn, s, grad s) dv.
\end{equation*}
The first variation of the energy $E$ is 
\begin{equation*}
\delta E = \int_V \left(\frac{\partial \psi}{\partial \bfn}\cdot \delta\bfn + \frac{\partial \psi}{\partial grad \bfn}:\delta(grad \bfn) + \frac{\partial \psi}{\partial s} \delta s + \frac{\partial \psi}{\partial grad s}\cdot \delta(grad s) \right)dv.
\end{equation*}
Integrate by parts the term involving $\delta(grad s)$ to obtain the $s$ evolution equation based on an $L^2$ gradient flow as 
\begin{equation}\label{eqn:s_evol}
\dot{s} =  \gamma \left[div \frac{\partial \psi}{\partial grad s} - \frac{\partial \psi}{\partial s}\right],
\end{equation}
where $\gamma$ is a dimensional constant. The result from the energy gradient flow method is equivalent to the evolution equation given in \eqref{eqn:s} for $m=2$.

\subsection{Phase transition model formulation} \label{sec:phase_model}
In Ericksen's model \cite{ericksen1991liquid}, the director field is represented by a unit length vector field $\bfn$. To practically implement the computation of a time-dependent nonlinear theory, we adopt an alternative vector field $\bfd$ to represent the director field subject to the constraint $|\bfd|^2=s^2$.

Assuming the generalized Parodi relation, the governing equations are an extension of the work in \cite{walkington2011numerical}, and take the form
\begin{eqnarray}\label{eqn:phase_tran}
\begin{aligned}
\rho \dot{v} + grad p - div \left( \frac{\partial R}{\partial grad \bfv} - (grad \bfd)^T \frac{\partial W}{\partial grad \bfd} - (grad s)\otimes \frac{\partial W}{\partial grad s}\right) = \rho \bff \\
\frac{\partial R}{\partial \dot{\bfd}} + \frac{\partial W}{\partial \bfd} - div \left( \frac{\partial W}{\partial grad \bfd} \right) + \lambda \bfd = \rho \bfm \\
\frac{\partial R}{\partial \dot{s}} + \frac{\partial W}{\partial s} -div \left( \frac{\partial W}{\partial grad s}\right) - \lambda s = \rho f_s 
\end{aligned}
\end{eqnarray}
where $\rho$ is the material density, $p$ and $\lambda$ are Lagrange multipliers dual to the constraints 
\[
div(v) = 0 
\quad \text{ and } \quad
|\bfd|^2 - s^2 = 0,
\]
$W$ is a modified Oseen-Frank energy, and $R$ is an appropriately designed dissipation function. 

We introduce the modified Oseen-Frank energy as 
\begin{eqnarray}\label{eqn:w_energy}
\begin{aligned}
W(\bfd, grad \bfd, s, grad s) = \frac{k_1}{2}div(\bfd)^2 + \frac{k_2}{2}( \bfd \cdot curl(\bfd))^2 \\
+ \frac{k_2-k_4}{2}(|grad\bfd|^2 - div(\bfd)^2 - |curl(\bfd)|^2) + \frac{k_3}{2}|\bfd \times curl(\bfd)|^2 \\
+ \frac{L_1}{2} |grad s|^2 + f(s) + g(grad s, \bfd),
\end{aligned}
\end{eqnarray}
where $k_1$, $k_2$, $k_3$ and $k_4$ correspond to the Frank constants, $L_1$ is the Leslie parameter and $f(s)$ is a non-convex function of s indicating the preferred phase state. The $(s, \bfd)$ modified Oseen-Frank energy function has been further augmented by the function $g(grad s,\bfd)$ which is a non-convex function representing interfacial energy. A natural candidate for $g(grad s, \bfd)$ is given as
\begin{equation}\label{eqn:g_energy}
g(grad s,\bfd) = |grad s|\left[\sigma_0 \left( 1+w\frac{(grad s \cdot \bfd)^2}{|grad s|^2|\bfd|^2}\right)\right],
\end{equation}
where $\sigma_0$ is an isotropic interfacial energy and $w$ is the anchor coefficient \cite{kim2013morphogenesis}. This is an adaption of the Rapini-Papoular function \cite{rapini1969distorsion}. The analog of the Parodi condition has Raleighian
\begin{eqnarray*}
R = (\gamma_0/2) (\bfd \cdot \bfD \bfd)^2 
+ (\hat{\gamma}_2/2) |\bfd \otimes \bfD \bfd|^2 \\
+ (\gamma_1/2) |\mathring{\bfd}|^2 
+ \gamma_2 \mathring{\bfd} \cdot \bfD \bfd
+ \beta_1 \dot{s} \bfd \cdot \bfD \bfd + (\beta_2/2) \dot{s}^2,
\end{eqnarray*}
where $\mathring{\bfd}:= \bfR^* \frac{d}{dt}(\bfR^{*T}\bfd) = \dot{\bfd}-\bfOmega \bfd$ is the convected derivate of $\bfd$ with respect to $\bfR^*$ (also called the Jaumann derivative), and $\bfR^*$ satisfies $\dot{\bfR^*}\bfR^{*T}=\bfOmega$. $\bfD$ and $\bfOmega$ are the symmetric and skew parts of the velocity gradient. The coefficients may depend upon $(s,\bfd,grad s, grad \bfd)$ and temperature. Equivalence between (\ref{eqn:s}) and the $s$ evolution embedded in (\ref{eqn:phase_tran}) is obtained by setting $\beta_1 = 0$ and $\beta_2 = B_m/|grad s|^{2-m}$, in which $R$ depends upon $grad s$.

However, since the non-convexity of interfacial energy involves $grad s$, it is possible that the evolution equation for $s$ is numerically unstable in the cases where $w$ is large. Recall the $s$ evolution equation in \eqref{eqn:s} is
\begin{equation*}
\dot{s}  = \frac{1}{B_m} |grad s|^{2-m} \rho \left[-\frac{\partial \psi}{\partial s} + div \left(\frac{\partial \psi}{\partial (grad s)}\right)\right],
\end{equation*}
where $\psi$ is taken as $W(\bfd, grad \bfd, s, grad s)$. Then with the energy density given in \eqref{eqn:w_energy}, $\frac{\partial \psi}{\partial grad s}$ is calculated as
\begin{eqnarray*}
\left(\frac{\partial \psi}{\partial grad s}\right)_{i} = L_1 (grad s)_{i} + \frac{\sigma_0}{|grad s|}(grad s)_{i} + \frac{2\sigma_0w}{|grad s||\bfd|^2}(d_id_js_{,j})-\frac{\sigma_0w cos^2\theta}{|grad s|}(grad s)_i \\
+\text{other terms},
\end{eqnarray*}
with $\theta$ being the angle between the interface normal direction and the tactoid interface, i.e. the angle between the directions $grad s$ and $\bfd$. Thus, after substituting $\frac{\partial \psi}{\partial grad s}$, we have
\begin{eqnarray*}
\dot{s} = C\left\{div\left[\left(\left(L_1+\frac{\sigma_0}{|grad s|}\right)\bfI 
-w\left(\frac{\sigma_0 cos^2\theta}{|grad s|}\bfI-\frac{2\sigma_0 }{|grad s||\bfd|^2}\bfd \otimes \bfd\right)\right)grad s \right]\right\}\\
+\text{other terms},
\end{eqnarray*}
where $C = \frac{|grad s|^{2-m} \rho}{B_m}$. Denote the diffusion tensor $\bfA$ as 
\[
\bfA = \left(L_1+\frac{\sigma_0}{|grad s|}\right)\bfI 
-w\left(\frac{\sigma_0 cos^2\theta}{|grad s|}\bfI-\frac{2\sigma_0 }{|grad s||\bfd|^2}\bfd \otimes \bfd\right).
\]
Then the $s$ evolution equation can be written as 
\begin{equation}\label{eqn:s_evol_W}
\dot{s} = C div (\bfA grad s) + \text{other terms}.
\end{equation}
Since $\bfd\cdot grad s$ is about $0$ near the tactoid interface where $grad s$ is nonzero (note that\eqref{eqn:g_energy} implies that $\bfd$ prefers to be perpendicular to $grad s$ to minimize interfacial energy), the diffusion tensor$\bfA$ in $div(\bfA grad s)$ may be negative-definite depending on the relative magnitude of $w$, a potential cause for numerical instability. 

In order to deal with this problem, we introduce a new field $\bfp$ representing the interfacial normal whose reciprocal magnitude roughly represents the width of the interface. The modified energy density with this new state descriptor is written as follows:
\begin{eqnarray*}
W(\bfd,\nabla \bfd, s, \nabla s, \bfp) = \frac{k_1}{2}div(d)^2 + \frac{k_2}{2}( \bfd \cdot curl(\bfd))^2 \\
+ \frac{k_2-k_4}{2}(|\nabla\bfd|^2 - div(\bfd)^2 - |curl(d)|^2) + \frac{k_3}{2}|\bfd \times curl(\bfd)|^2 \\
+ \frac{L_1}{2} |grad s- \bfp|^2 + f(s) + g(\bfp, \bfd),
\end{eqnarray*}
where $f(s)$ is still the non-convex function of $s$ in \eqref{eqn:w_energy} and $g(\bfp,\bfd)$ is a modified non-convex function representing interfacial energy given as 
\begin{equation}\label{eqn:interface_energy}
g(\bfp,\bfd) = |\bfp|\left[\sigma_0\left(1+w\frac{(\bfp \cdot \bfd)^2}{|\bfp|^2|\bfd|^2}\right)\right].
\end{equation}

By placing the non-convexity of the interfacial energy to be a function of $\bfp$ and $\bfd$, and elastically penalizing the difference between $\bfp$ and $grad s$, we get a stable system for the phase transition model. With the modified energy density with the new state descriptor, the dissipation in \eqref{eqn:s_dissipation} (we ignore material motion for simplicity) can be written as
\begin{eqnarray*}
&\int_V[\bfLambda :\bfM] - \rho \dot{\psi}] dv \ge 0 \\
&\Rightarrow \int_V \left[ \Lambda_{ij}\omega_{i,j} -\rho \frac{\partial W}{\partial d_i} \dot{d_i} - \rho \frac{\partial W}{\partial (d_{i,j})} \dot{\overline{d_{i,j}}} - \rho \frac{\partial W}{\partial s}\dot{s} - \rho \frac{\partial W}{\partial (s_{,j})}\dot{\overline{s_{,j}}} - \rho \frac{\partial W}{\partial p_i} \dot{p_i} \right]dv \ge 0.
\end{eqnarray*}
Following the same procedure as in Sec. \ref{sec:evols}, we can verify that the dissipation is non-negative when $\dot{\bfp}$ is in the direction of $-\frac{\partial W}{\partial \bfp}$. Thus, the dynamic evolution equation of the $\bfp$ field is given as
\begin{equation}\label{eqn:p_evol_W}
\dot{\bfp} = -Q\frac{\partial W}{\partial \bfp} = -Q\left[L_1(\bfp-grad s)+\frac{\partial g}{\partial \bfp} \right],
\end{equation}
where $Q$ is a material dependent constant. An example of the advantage of the modified $\bfp$ model is discussed in Section \ref{sec:results}.

 \textit{The variables $\bfd$, $s$, and the anchoring coefficient $w$ are dimensionless. The variable $\bfp$ has dimension $[\bfp] = Length^{-1}$. The physical dimensions of the parameters in the modified Oseen-Frank energy are $[k_1] = Force$, $[k_2] = Force$, $[k_3] = Force$, $[k_4] = Force$, $[L_1] = Force$, and $[\sigma_0]=Force \times Length^{-1}$. The physical dimensions of the coefficients $C$ in \eqref{eqn:s_evol_W} and $Q$ in \eqref{eqn:p_evol_W} are $[C] = Length^2\times Time^{-1} \times Force^{-1}$, and $[Q] = Time^{-1} \times Force^{-1}$.}

To non-dimensionalize the above parameters, we introduce the following dimensionless variables,
\begin{eqnarray*}
\tilde{\bfp} = R\bfp; \quad \tilde{k}_i = \frac{k_i}{k_1};\quad \tilde{L}_1 = \frac{L_1}{k_1}; \quad \tilde{\sigma}_0 = R\frac{\sigma_0}{k_1}; \quad \tilde{l} = \frac{l}{R},
\end{eqnarray*}
where $l$ is the dimensional length, $\tilde{l}$ is the dimensionless length, and $R$ is half of a typical tactoid size. In this work, we assume $k_1=k_2=k_3 = k$ (except in Sec. \ref{sec:para_frank}), $k_4=0$, and $L_1=k$. Therefore, $\tilde{k}_1 = \tilde{k}_2 = \tilde{k_3} = 1$, $\tilde{k_4} = 0$, and $\tilde{L}_1 = 1$. The dimensionless $\tilde{\sigma}_0$ physically represents the ratio of the total surface energy and the total elastic energy, which would be $\frac{\sigma_0 R^2}{k R}$ for a three-dimensional nematic tactoid \cite{lavrentovich1998topological}. In this work, we assume $R$ to be $10$ $\mu m$, based on the estimate of the long-axis length of a `two-cusp tactoid' of $20$ $\mu m$ given in \cite{kim2013morphogenesis}. The physical parameters of LCLCs are adopted from \cite{kim2013morphogenesis} as follows: $k=2\times 10^{-12}N$, $\sigma_0=10^{-4} J/m^2$, which implies $\tilde{\sigma}_0 = 500$. Since we do not focus on the evolution rates of $s$ and $\bfp$, we assume that the time scales in $s$ and $\bfp$ evolutions are similar by setting $Q=\frac{C}{R^2}$.

\section{Tactoid equilibrium and phase transition results} \label{sec:results}

We explore the capability of the phase transition model proposed in Sections \ref{sec:evols} and \ref{sec:phase_model} by solving tactoid equilibrium and dynamic problems. In static problems, both the initialized shapes from the Wulff construction and arbitrary initialized shapes are discussed. In addition, the nematic-isotropic phase transition and the formation of disclinations are also studied. 

\subsection{Tactoid static equilibrium}

\begin{figure}
\centering
\subfigure[The initialized tactoid shape and director field with $w=0.1$.]{
\includegraphics[width=0.45\textwidth]{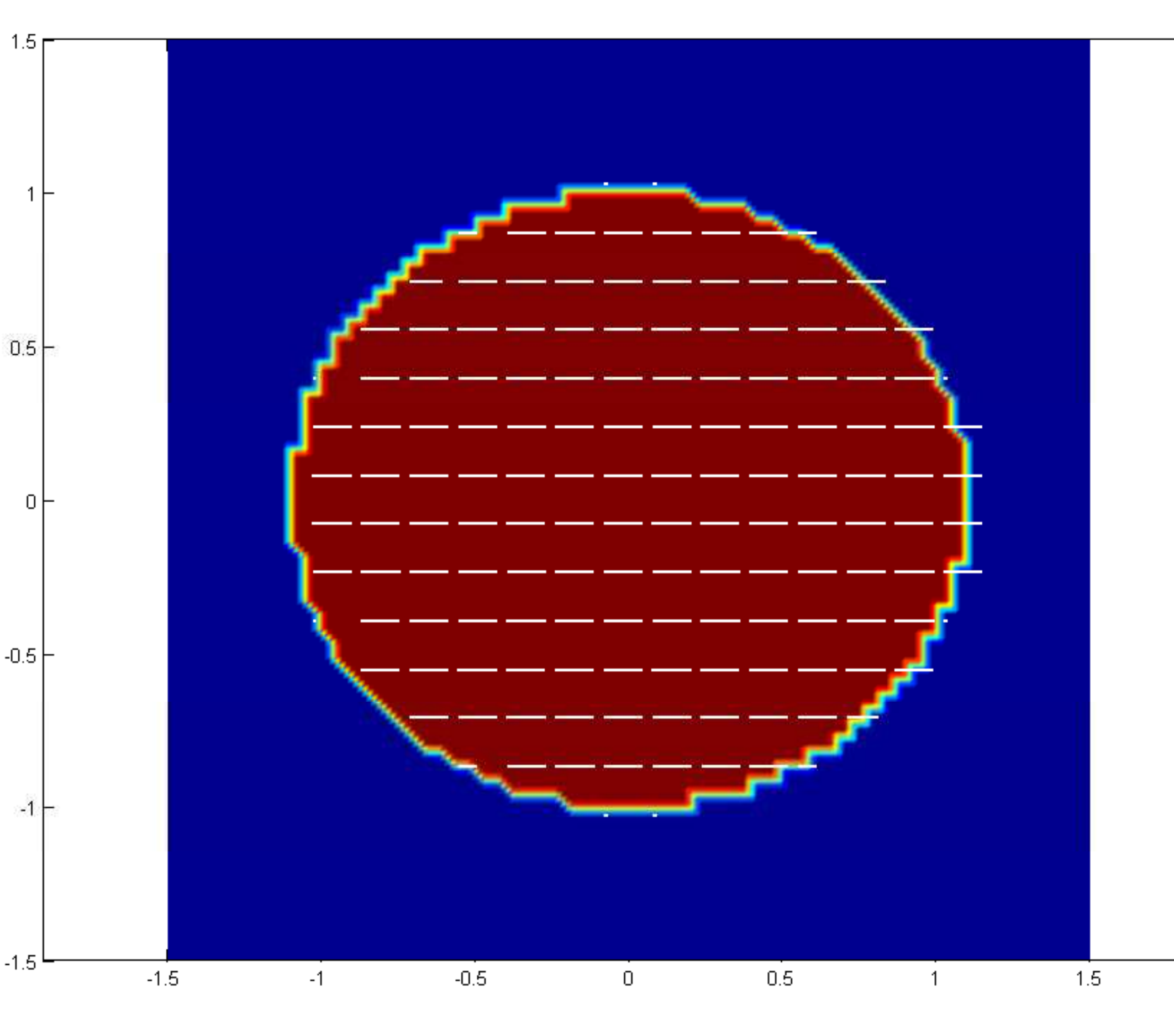}
}\qquad
\subfigure[The equilibrium of the tactoid shape and director field with $w=0.1$.]{
\includegraphics[width=0.45\textwidth]{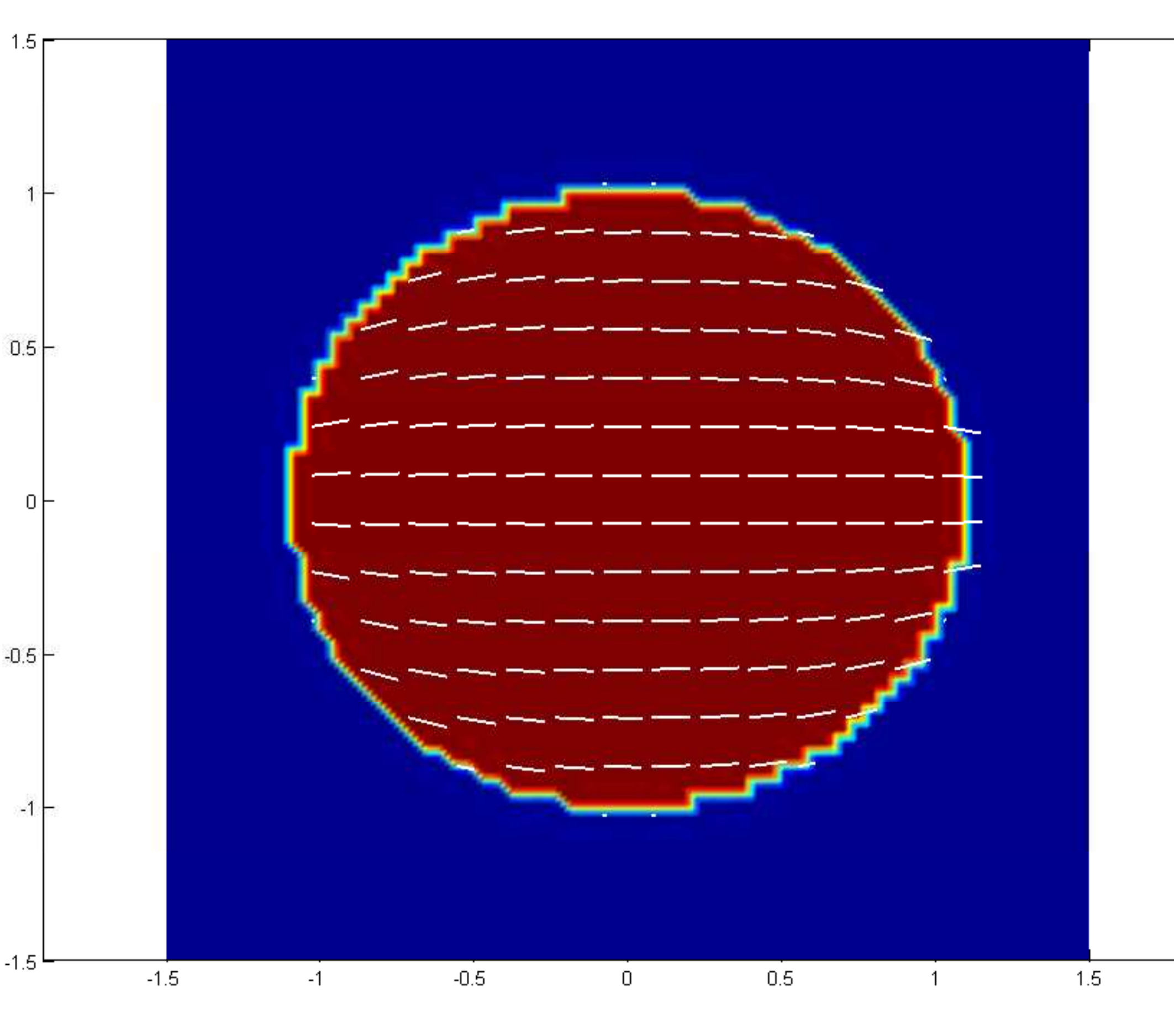}
}
\subfigure[The initialized tactoid shape and director field with $w=1$.]{
\includegraphics[width=0.45\textwidth]{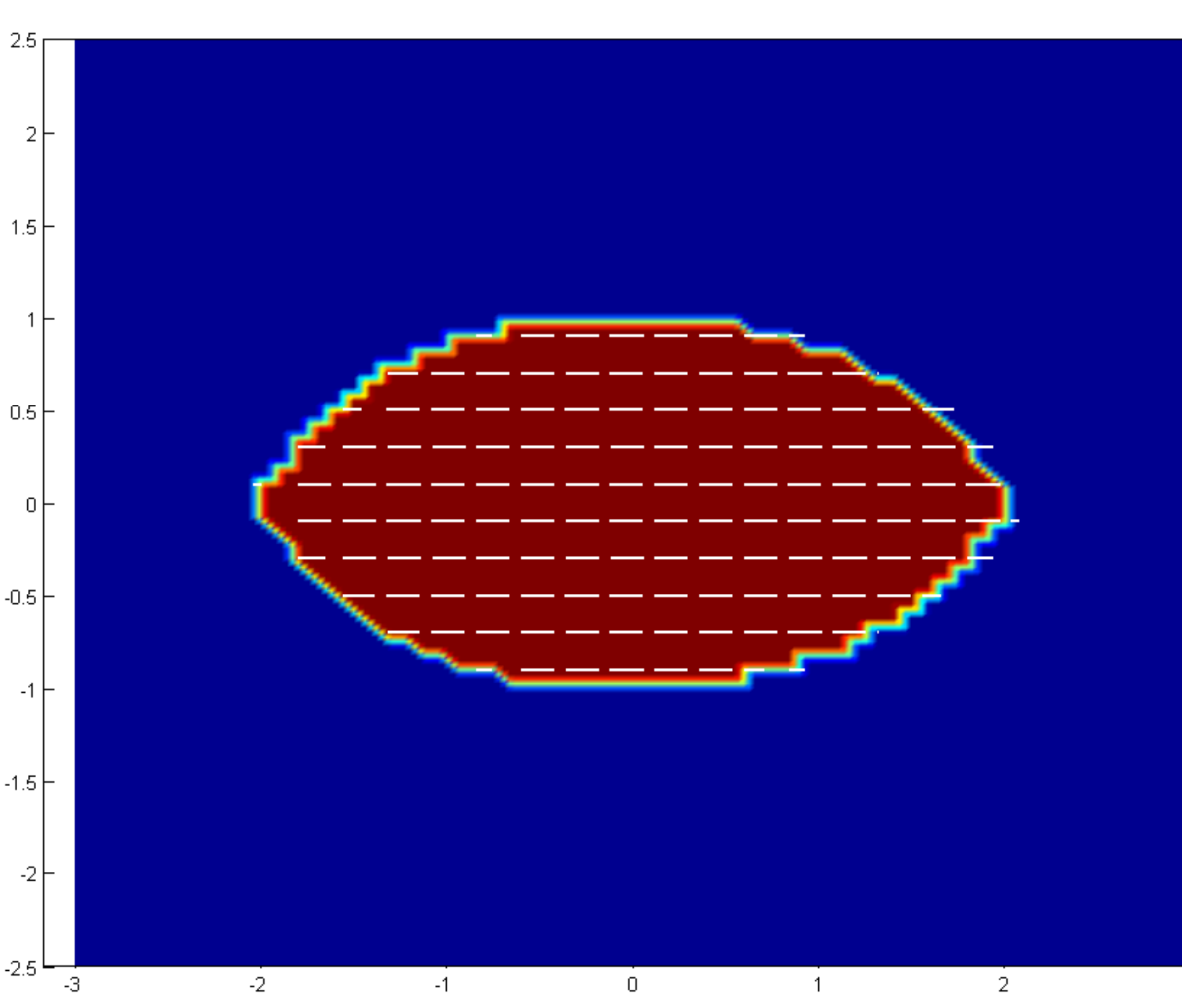}
}\qquad
\subfigure[The equilibrium of the tactoid shape director field with $w=1$.]{
\includegraphics[width=0.45\textwidth]{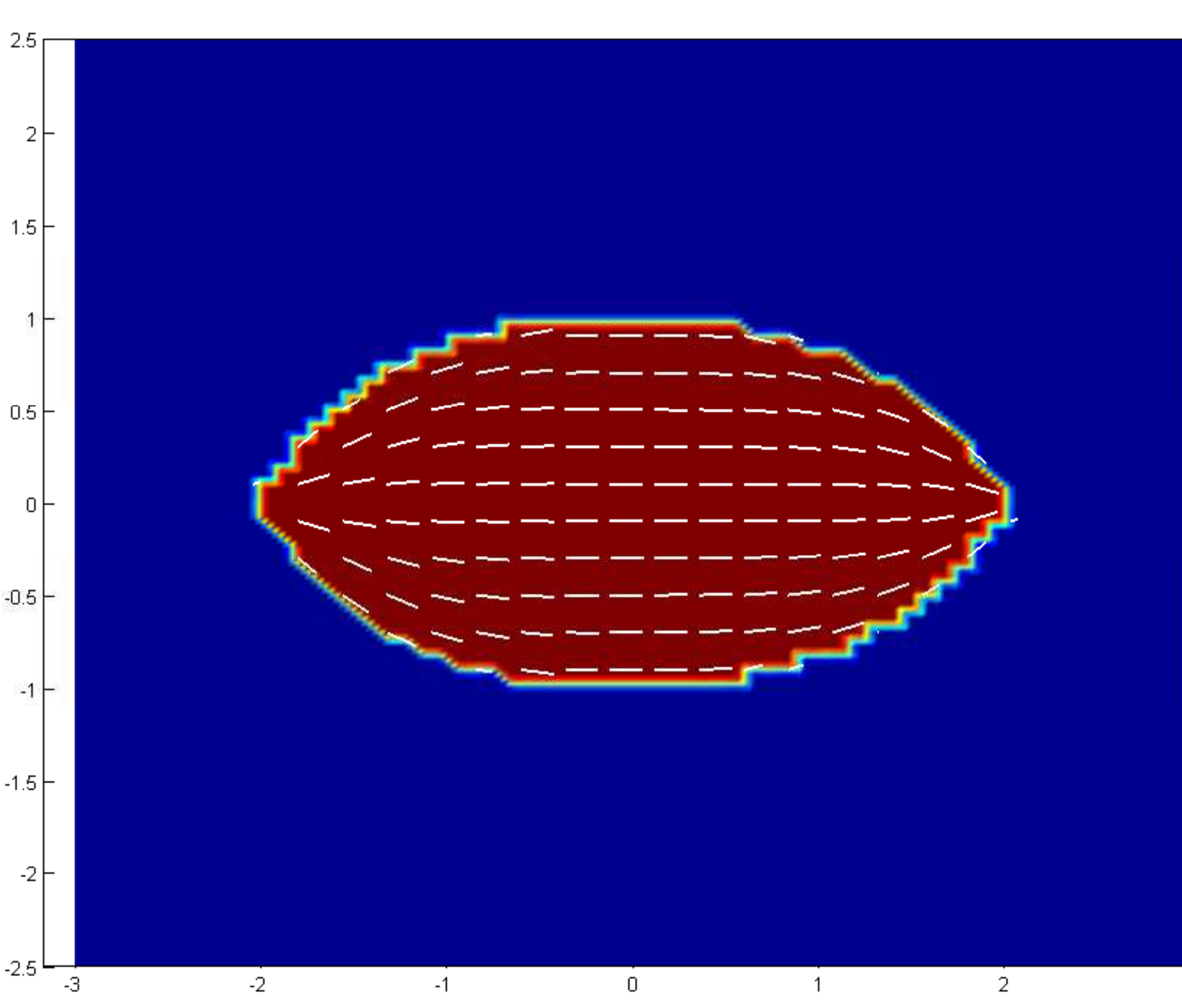}
}
\subfigure[The initialized tactoid shape and director field with $w=2$.]{
\includegraphics[width=0.45\textwidth]{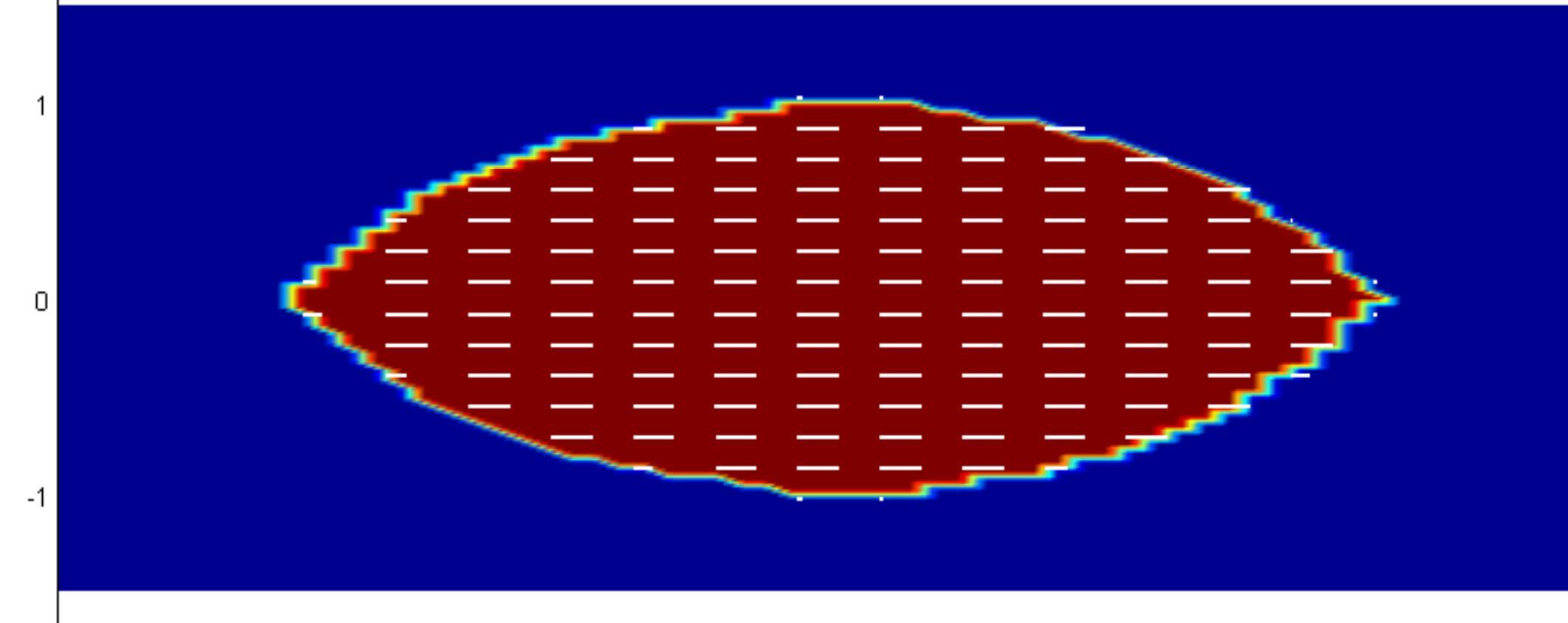}
}\qquad
\subfigure[The equilibrium of the tactoid shape and director field with $w=2$.]{
\includegraphics[width=0.45\textwidth]{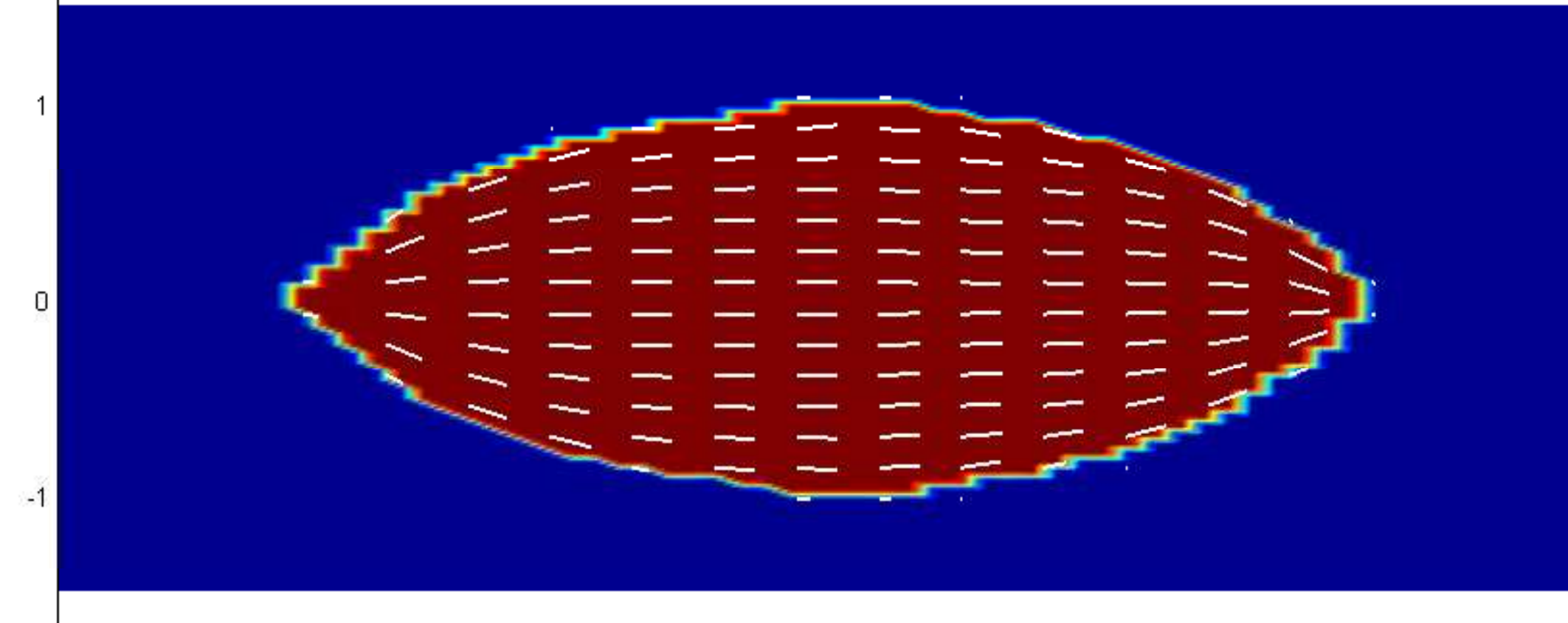}
\label{fig:results_static_a_3}
}
\caption{Initializations and equilibria of tactoid static problems with different anchor coefficients. The red color represents $s=1$, the blue color represents $s=0$ and the white dash lines represent the director field. The tactoid initializations are calculated from the Wulff construction.}
\label{fig:results_static_a}
\end{figure}

We discuss the results of tactoid equilibrium calculations with different anchor coefficients $w$. Based on the Wulff construction of equilibrium shapes of perfect crystals with the interfacial energy given in  \eqref{eqn:interface_energy}, we can construct the equilibrium shapes of tactoids under the condition of constant surface area and a frozen director field \cite{kim2013morphogenesis, burton1951growth, landau1964theoretical, oswald2005nematic, wheeler2006phase, khare2003determining, nozieres1992shape}. In the static problem, we assume the non-convex function $f(s)$ in the energy density has identical values at $s=0$ and $s=1$ characterizing its minimum. Fig. \ref{fig:results_static_a} shows the initializations and the corresponding equilibrium results for various tactoids. The tactoid is initialized in the nematic $s = 1$ state and the matirx in the isotropic $s = 0$ phase. For fixed $w$, no large scale evolution is seen to occur in tactoid shapes, but director re-orientation occurs as the system seeks out a local minima.

The left column in Fig. \ref{fig:results_static_a} shows the initializations of the director field and tactoid shapes for different anchor coefficients $w$. The initialized tactoid shapes are calculated from the Wulff construction and the director fields start from a uniform unit vector field where $s=1$. The right column in Fig. \ref{fig:results_static_a} are the equilibrium configurations corresponding to the initializations. It shows that with increasing $w$, the single tactoid shape started from the Wulff construction transforms from sphere-like to ellipse-like shape. In all cases, given the interfacial energy in \eqref{eqn:interface_energy}, the director field tends to be perpendicular to the interface normal $grad s$. 

Recall that we introduced a new field $\bfp$ and discussed the theoretical motivation behind it in Section \ref{sec:phase_model}.  In Fig. \ref{fig:results_static_a_3}, the anchor coefficient $w$ is set to be large, $w=2$. In this case, without  introducing the $\bfp$ field, the computation is unstable and an equilibrium could not be found. With the introduced field $\bfp$, this case can be solved with result shown in Fig. \ref{fig:results_static_a_3}. The results of various tactoid shapes show that cusps are recovered in our model, matching with experimental observations \cite{kim2013morphogenesis}.

\begin{figure}
\centering
\subfigure[The spherical initialized tactoid shape with $w=2.5$.]{
\includegraphics[width=0.45\textwidth]{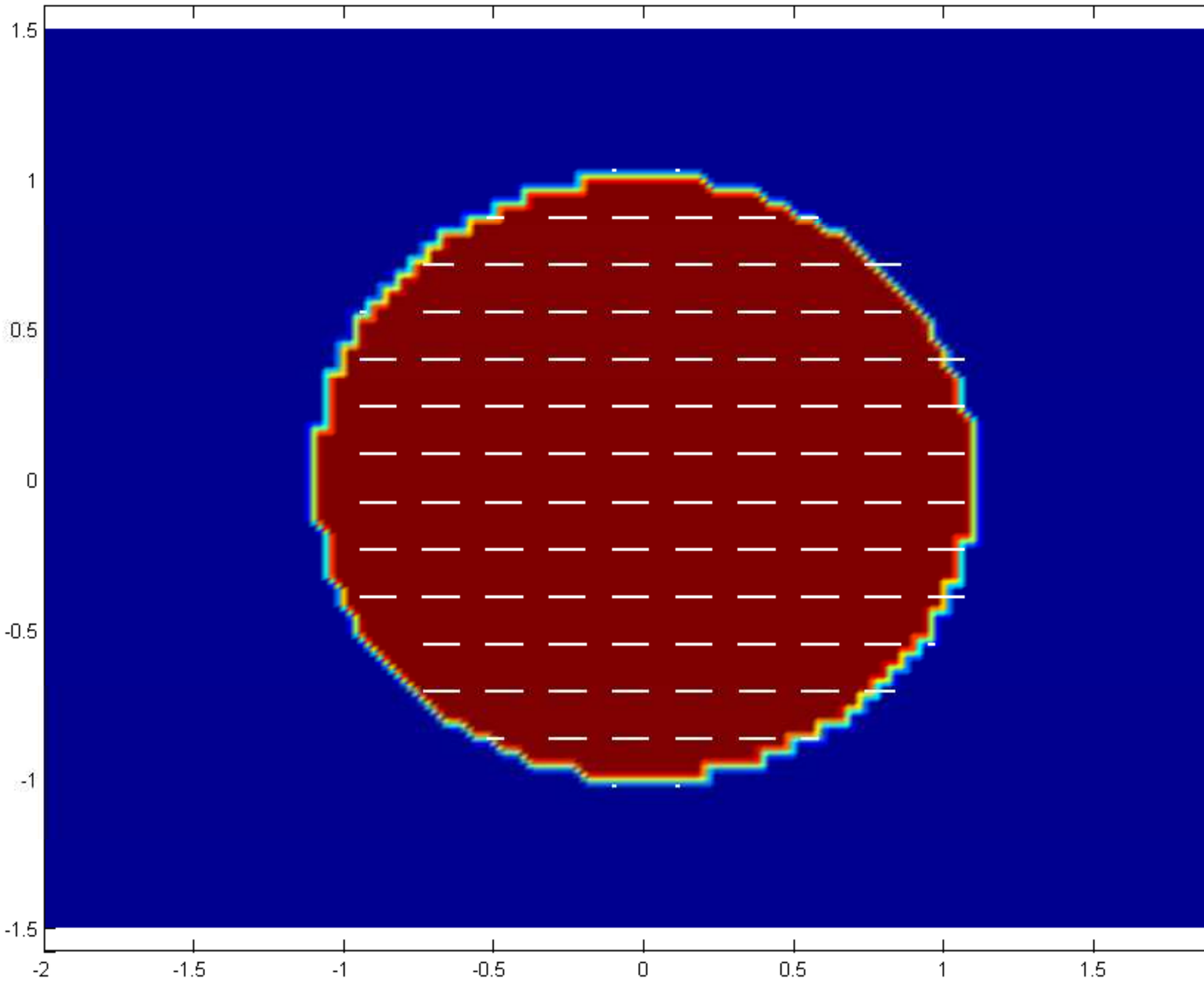}
\label{fig:result_static_b_1}
}\qquad
\subfigure[The equilibrium of the tactoid shape and director field with $w=2.5$. ]{
\includegraphics[width=0.45\textwidth]{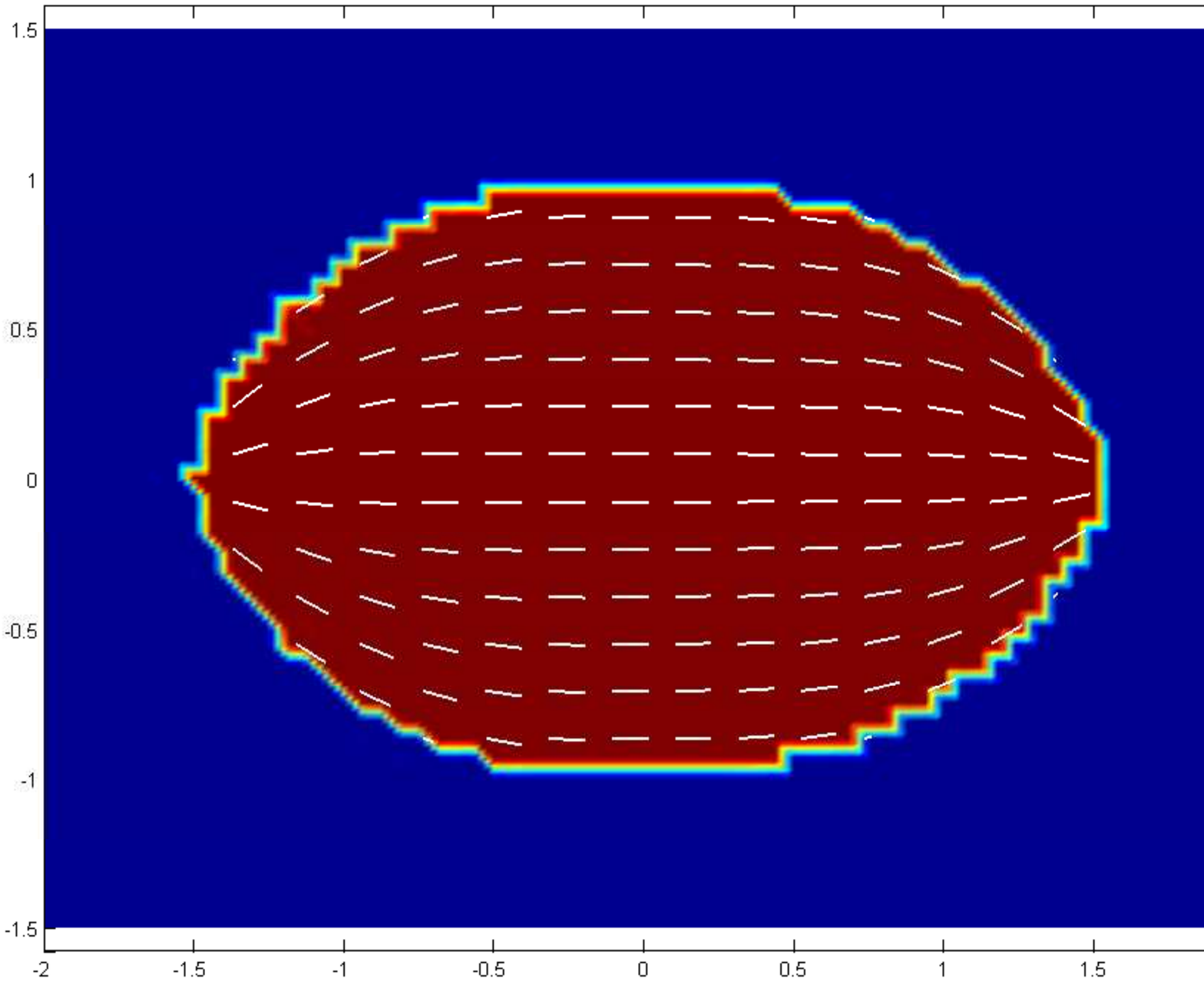}
\label{fig:result_static_b_2}
}
\caption{The initialized tactoid shape is a sphere with $w=2.5$. At the equilibrium, the spherical tactoid transforms to an ellipse-like tactoid and the director field evolves.}
\label{fig:result_static_b}
\end {figure}

The initialized tactoid shapes in Fig. \ref{fig:results_static_a} are based on the Wulff construction. The determined shape from the Wulff procedure depends on the value of $w$. In addition, the calculation shown in Fig. \ref{fig:result_static_b} explores the capability of the proposed model with a specified $w$ and an arbitrary initialized shape. In Fig. \ref{fig:result_static_b}, $w$ is assumed to be $2.5$ but the initialized tactoid shape is a sphere which clearly does not match with the Wulff construction. Fig. \ref{fig:result_static_b_1} is the initialization of the tactoid shape and the director field and Fig. \ref{fig:result_static_b_2} is the corresponding computed equilibrium state. It shows that the initialized spherical tactoid shape transforms to an elliptic shape due to the high value of $w$. 

\begin{figure}
\centering
\subfigure[The spherical initialized tactoid shape with $w=1.5$ and the initialized director field corresponds to a negative disclination of strength $-1$.]{
\includegraphics[width=0.45\textwidth]{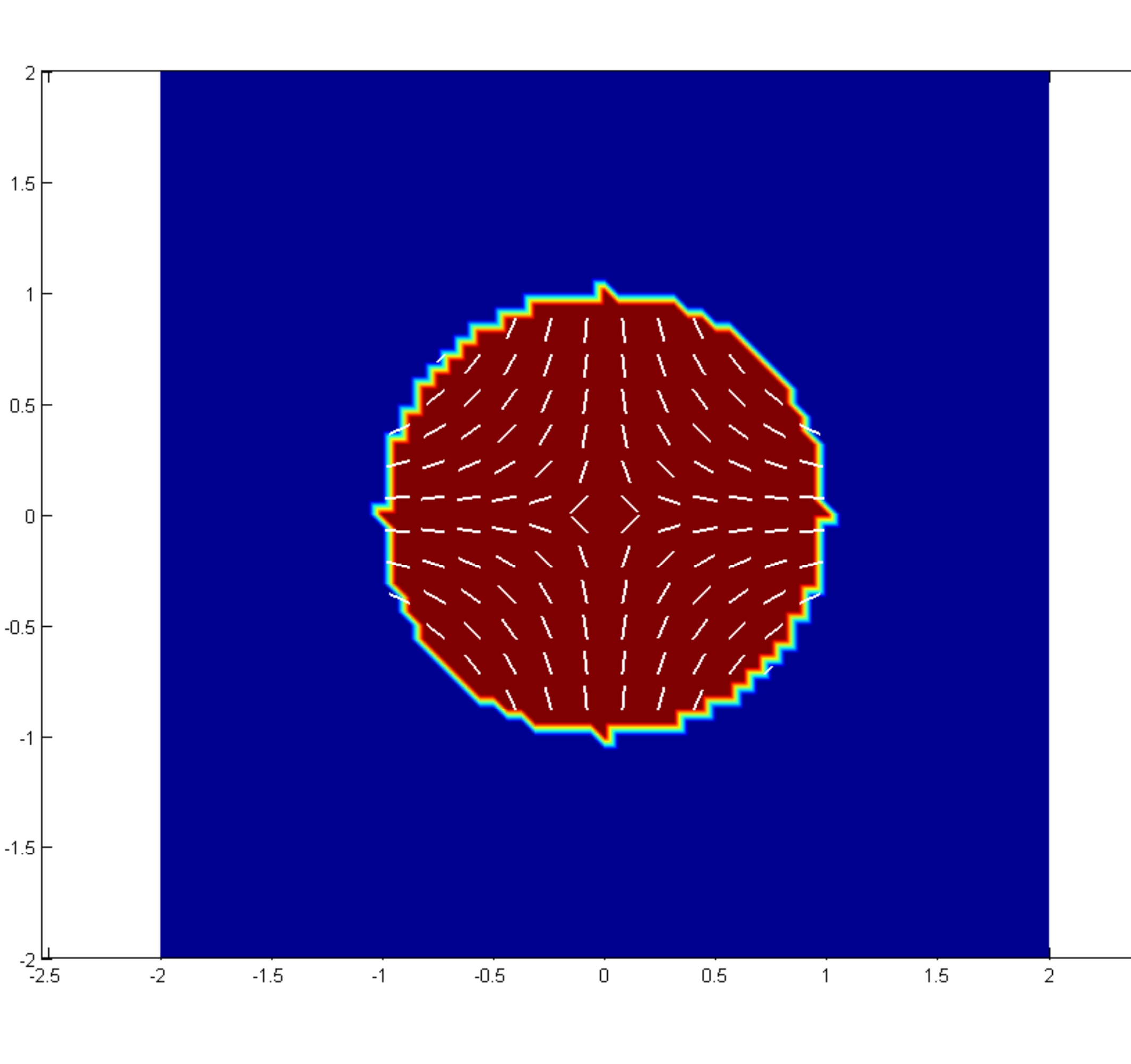}
\label{fig:result_static_c_1}
}\qquad
\subfigure[The tactoid shape and director field at equilibrium.]{
\includegraphics[width=0.45\textwidth]{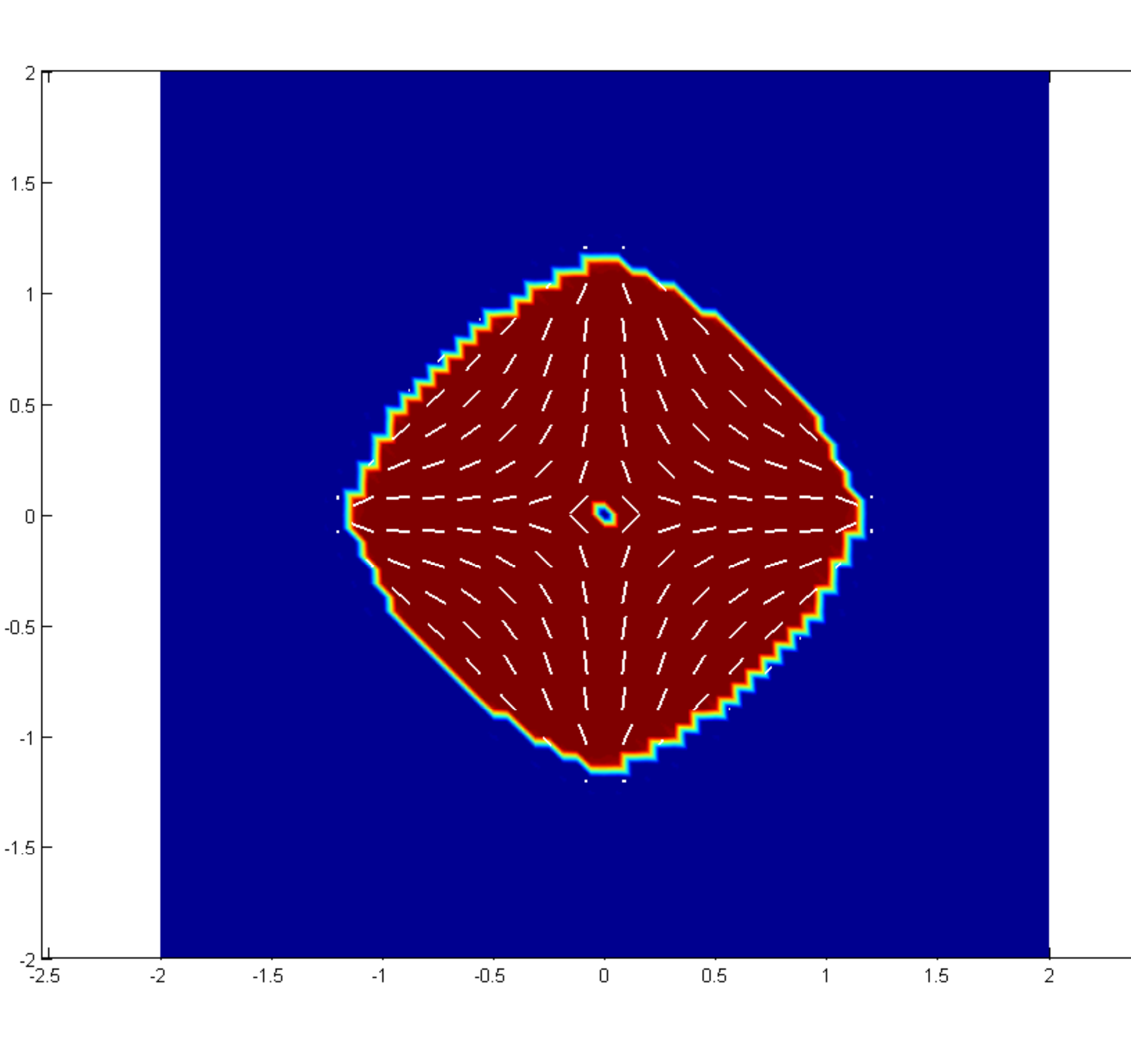}
\label{fig:result_static_c_2}
}
\caption{A spherical tactoid transforms to a rounded-square tractoid with $w=1.5$ and a negative disclination of strength $-1$. }
\label{fig:result_static_c}
\end {figure}

Fig. \ref{fig:result_static_c} shows another example with a non-Wulff constructed initialized shape in which $w=1.5$ and the director field is prescribed with a singularity corresponding to a negative disclination of strength $-1$. Fig. \ref{fig:result_static_c_1} is the initialized spherical tactoid shape and the director field with the discontinuity at the center of the tactoid. Fig. \ref{fig:result_static_c_2} shows the final equilibrium state indicating that the tactoid transforms to a rounded square, and a negative disclination (with its core in the isotropic phase $s=0$) exists at the center of the tactoid. 

\subsection{Dynamics of tactoids interaction}

\begin{figure}
\centering
\subfigure[The initialized tactoid shape and director field for two tactoids interaction.]{
\includegraphics[width=0.45\textwidth]{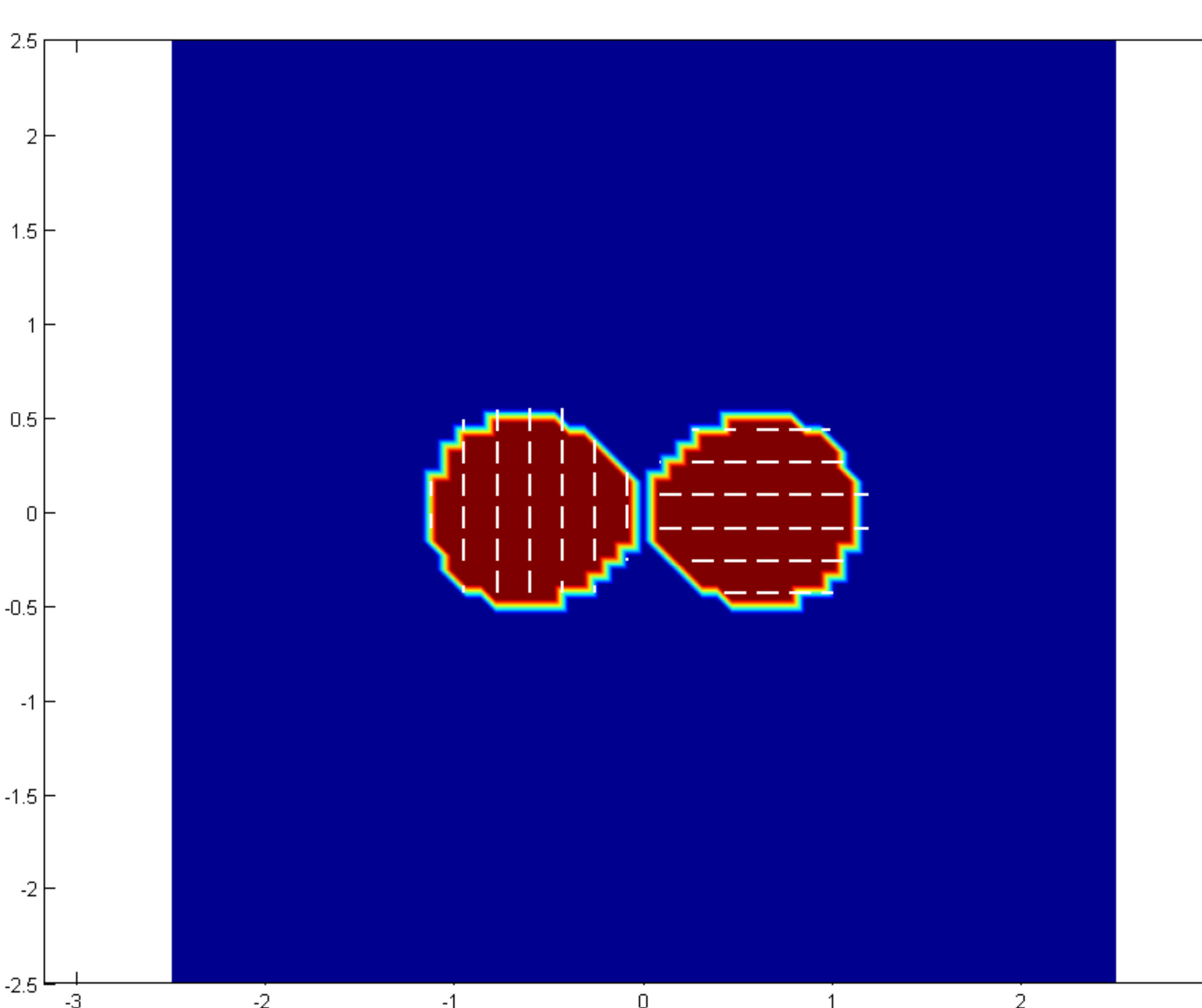}
\label{fig:results_interaction_1}
}
\subfigure[Two tactoids begin to merge and the director evolves.]{
\includegraphics[width=0.45\textwidth]{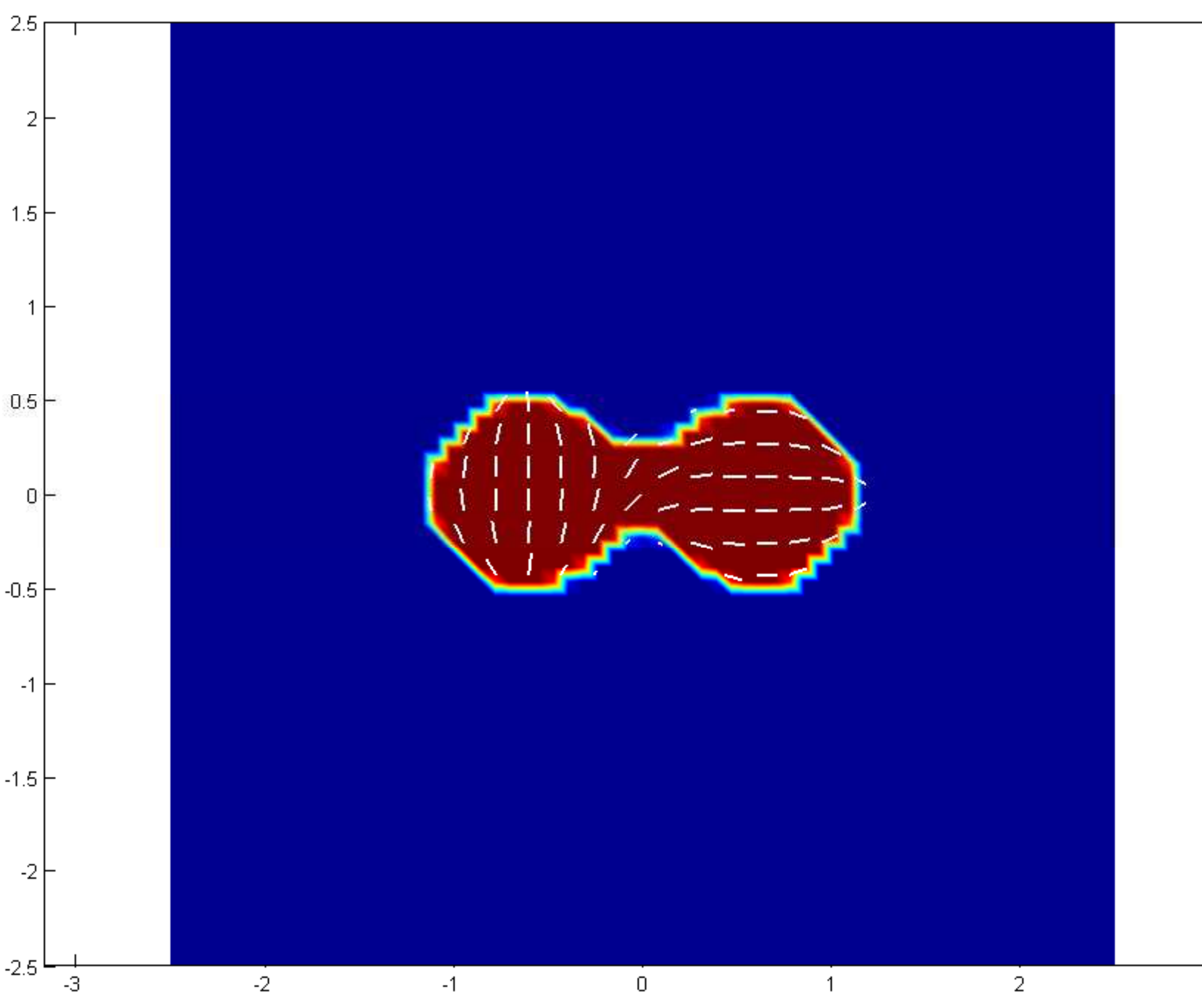}
\label{fig:results_interaction_2}
}
\subfigure[The director keeps evolving.]{
\includegraphics[width=0.45\textwidth]{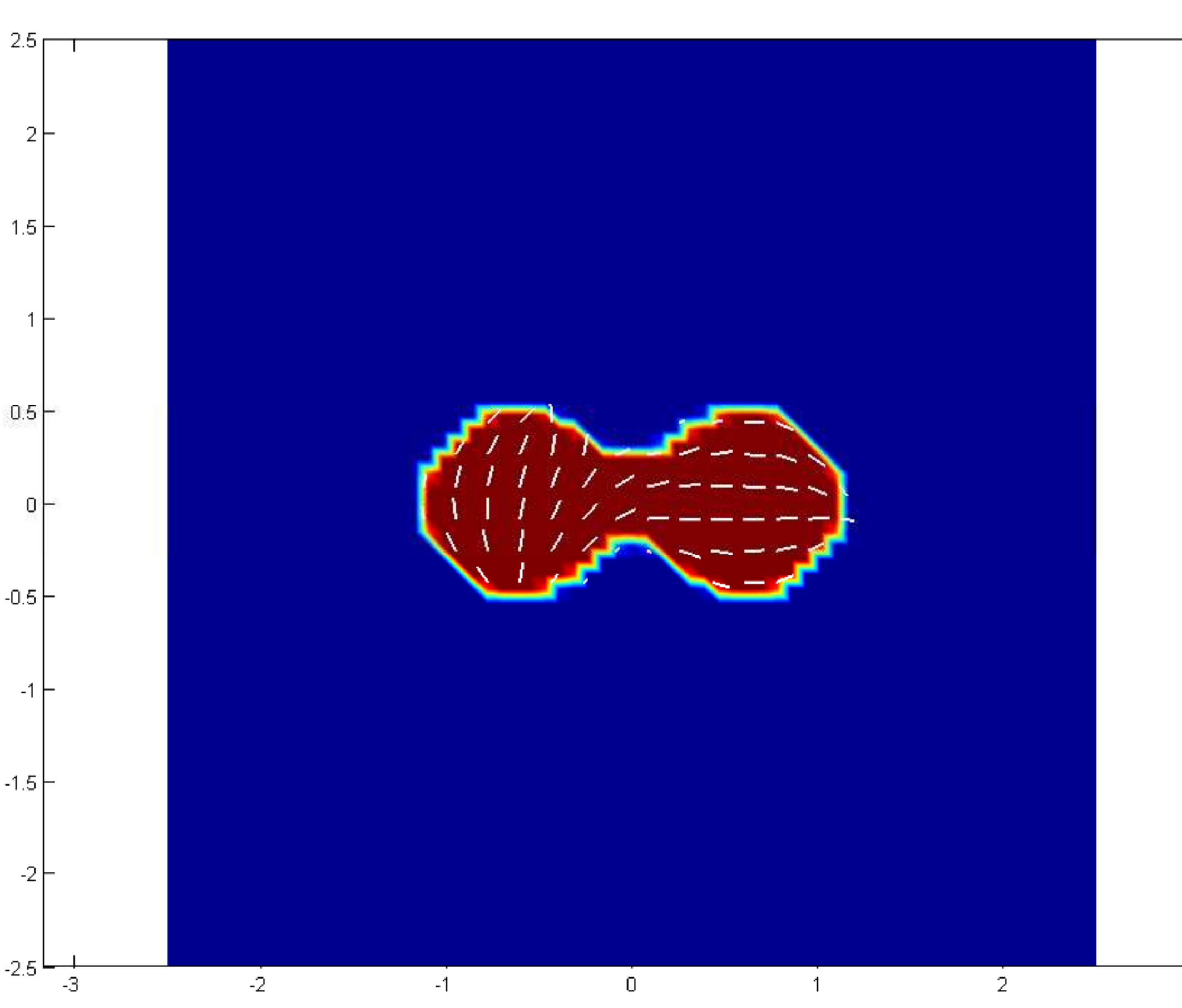}
\label{fig:results_interaction_3}
}
\subfigure[The equilibrium of two tactoid interaction.]{
\includegraphics[width=0.45\textwidth]{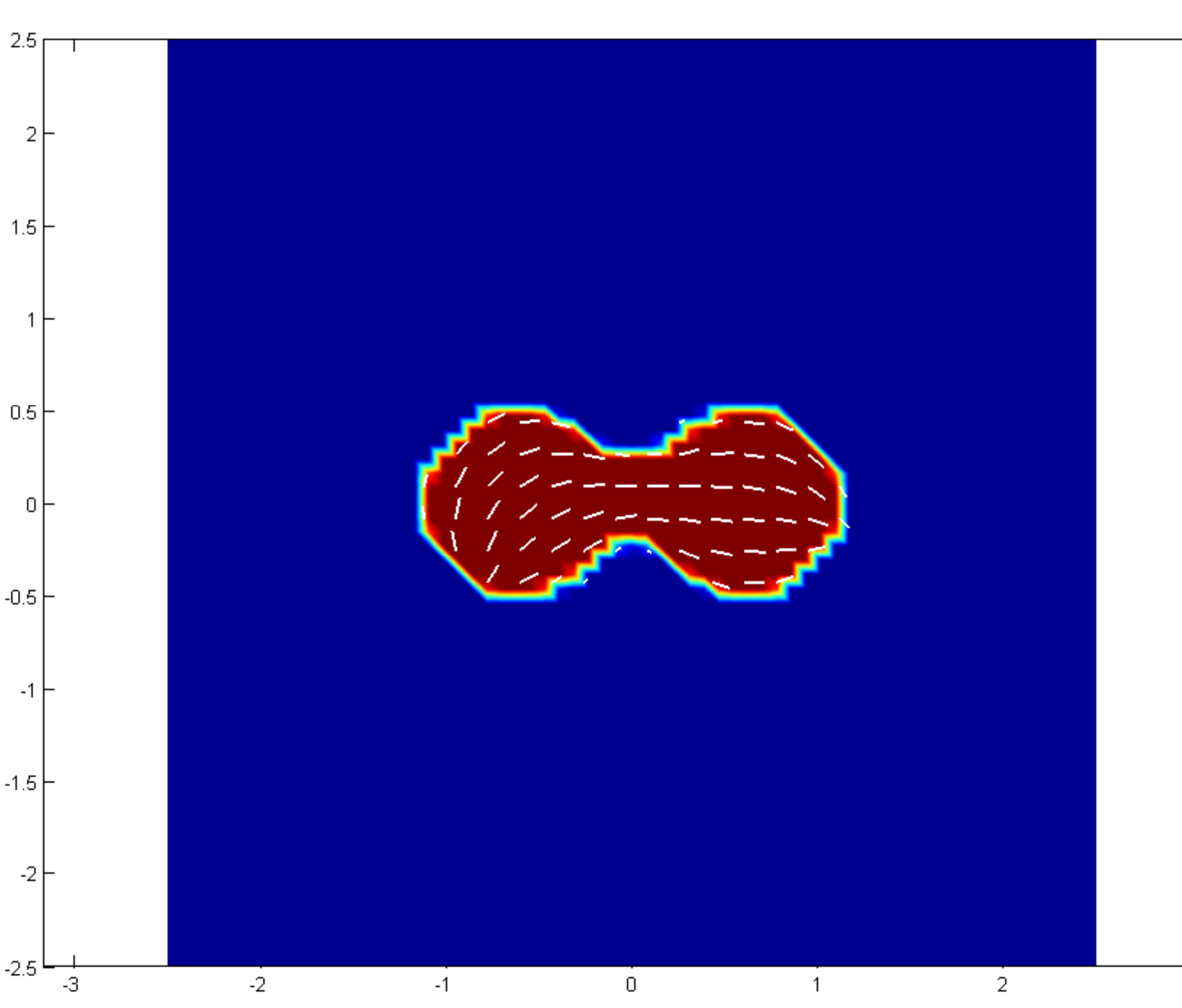}
\label{fig:results_interaction_4}
}
\caption{Interaction between two tactoids. These two tactoids tend to merge and the director field evolves.}
\label{fig:results_interaction}
\end {figure}

The interaction between two tactoids located close to each other is computed. Two spherical tactoids are initialized with different director orientations, as shown in Fig. \ref{fig:results_interaction_1}. Since these two tactoids are located very close to each other, they are expected to interact with each other. As the calculation progresses, the tactoids begin to merge and the director field evolves to minimize the total energy, as shown in Fig. \ref{fig:results_interaction}.

\begin{figure}
\centering
\includegraphics[width=0.5\textwidth]{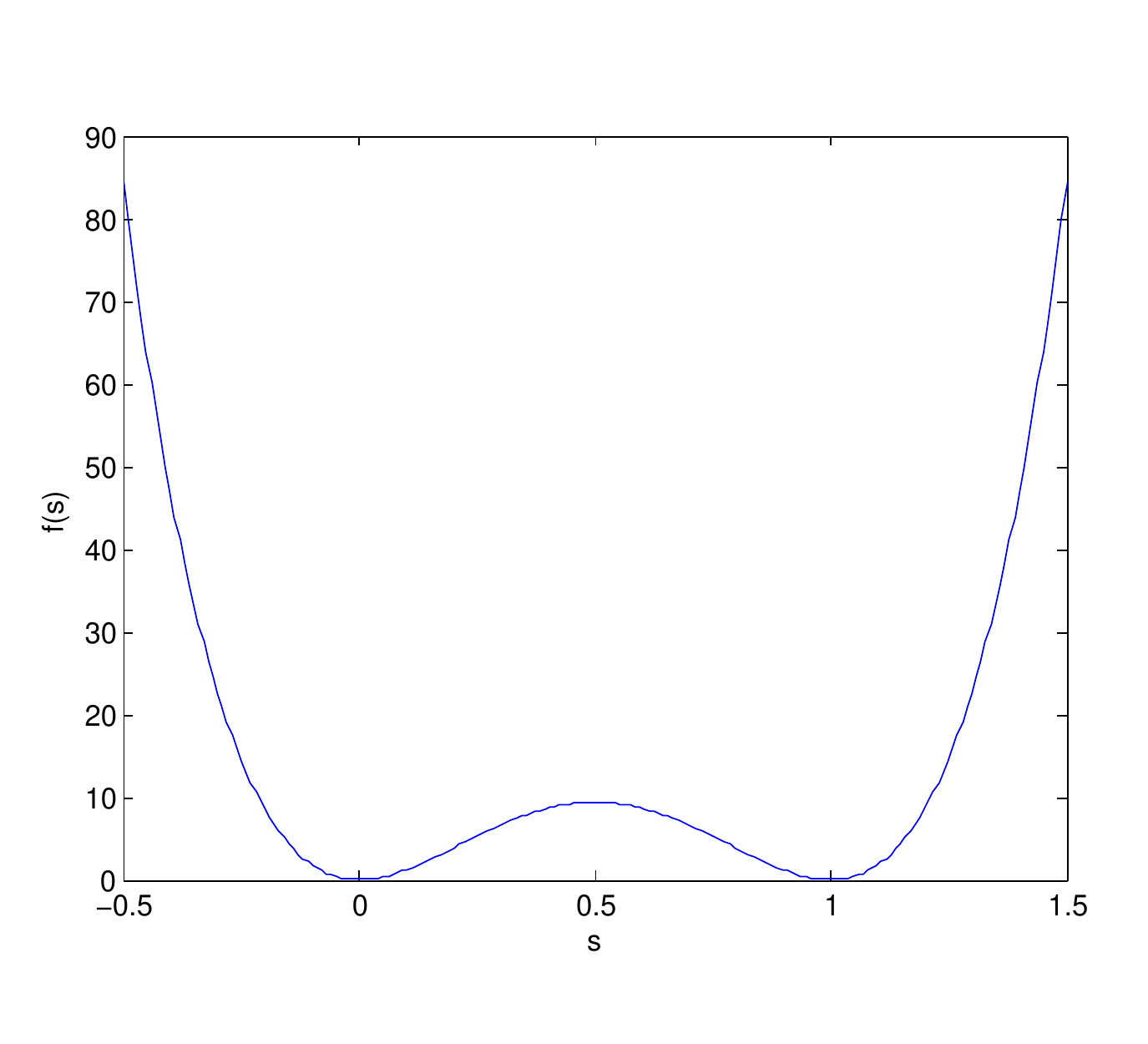}
\caption{The shape of $f(s)$ used in the two tactoid interaction calculation. The barrier between two wells at $s=0$ and $s=1$ is low.}
\label{fig:nonconvex}
\end {figure}

In this calculation, $m=0$ and the barrier of the non-convex function $f(s)$ in the energy density between $s=0$ and $s=1$ is low. The shape of the non-convex function $f(s)$ is shown in Fig. \ref{fig:nonconvex}. In the tactoid evolution, the effect of $m$ is critical. 
\begin{itemize}
\item For the static equilibrium problem of a single tactoid, with higher barrier of $f(s)$, a single tactoid will evolve to its equilibrium state with no problem. 
\item For the static equilibrium problem, with a low barrier of $f(s)$, and $m=2$, the single tactoid will diffuse into the isotropic matrix and the interface cannot maintain its shape. On the other hand, with a low barrier of $f(s)$, and $m=0$, the single tactoid will evolve to its equilibrium state. 
\item For dynamic problems, such as the tactoid interaction discussed in this section, the tactoids are not able to merge with a high barrier in $f(s)$. 
\item With a low barrier of $f(s)$ as applied in this calculation and $m=0$, the tactoids are able to move, expand, or merge.
\end{itemize}

To understand the reason for the effect of the energy barrier and $m$ value, recall that the $s$ evolution equation is given as
\begin{equation*} 
\dot{s}  = \frac{1}{B_m} |grad s|^{2-m} \rho [-\frac{\partial \psi}{\partial s} + div (\frac{\partial \psi}{\partial (grad s)})].
\end{equation*}
In the case of high barrier of $f(s)$, regardless of $m$, $s$ can barely evolve from their well values because of the high value of the `resisting force' from $\frac{\partial \psi}{\partial s}$. In the case of low barrier of $f(s)$, with $m=2$, there is no impediment for $s$ to evolve out of the isotropic well. In the case of low barier and with $m=0$, although the barrier of $f(s)$ is low, $s$ cannot evolve where $grad s$ is $0$.

This is analogous to a problem in \cite{zhang2016non}, where the dissipative dynamic behavior of disclinations in nematic liquid crystals is studied. By observing the effect of $m$ on low barrier cases, we show that the dynamic model based on \emph{kinematics} and thermodynamics is important for modeling dissipative dynamics.

\subsection{Phase transition}

\begin{figure}
\centering
\subfigure[The initialized tactoid shape and director field. Three spherical tactoids with same director fields are initialized.]{
\includegraphics[width=0.45\textwidth]{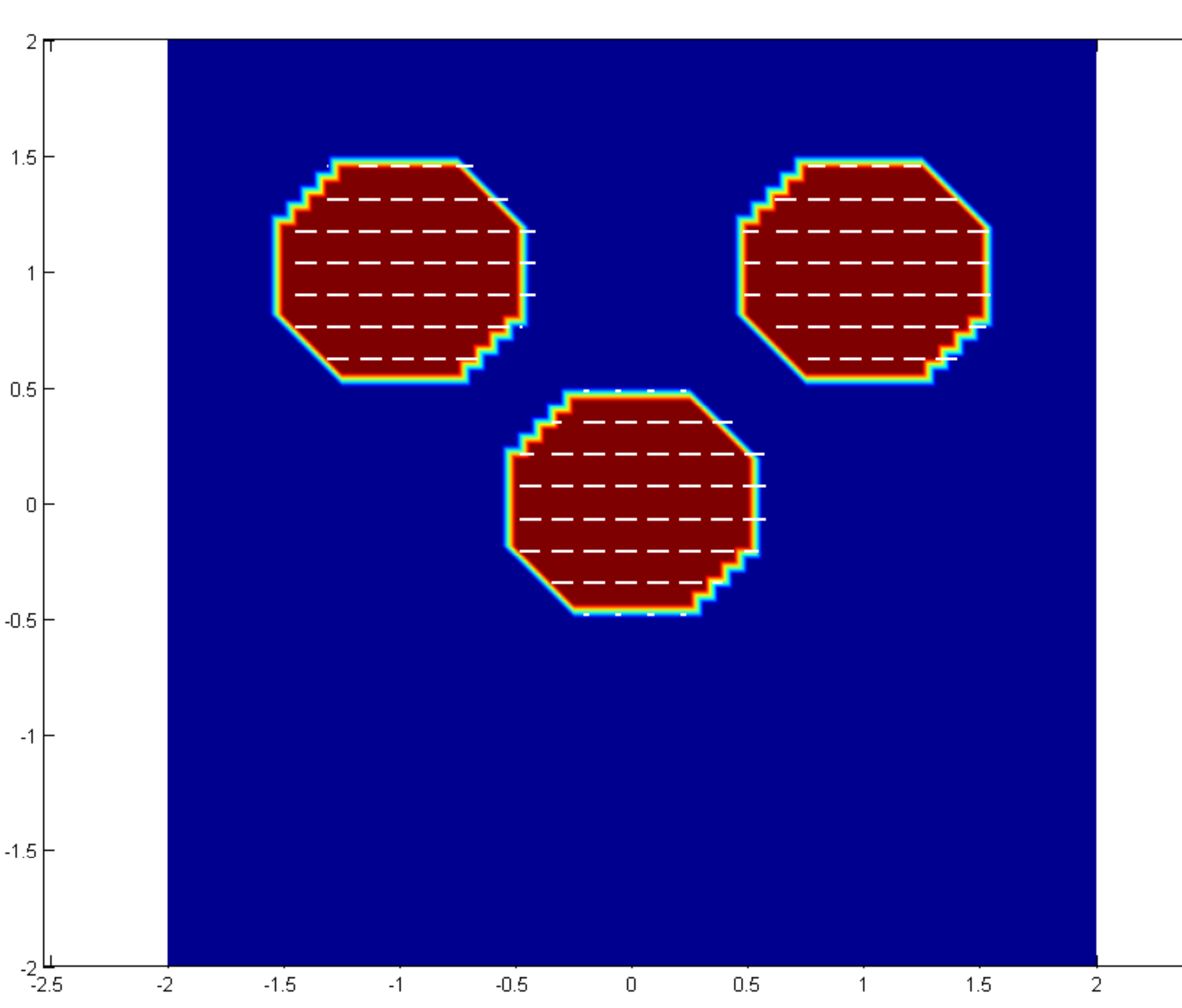}
\label{fig:results_phase_1}
} 
\subfigure[The tactoid shapes and director field at $t=0.1$. Three tactoids expand.]{
\includegraphics[width=0.45\textwidth]{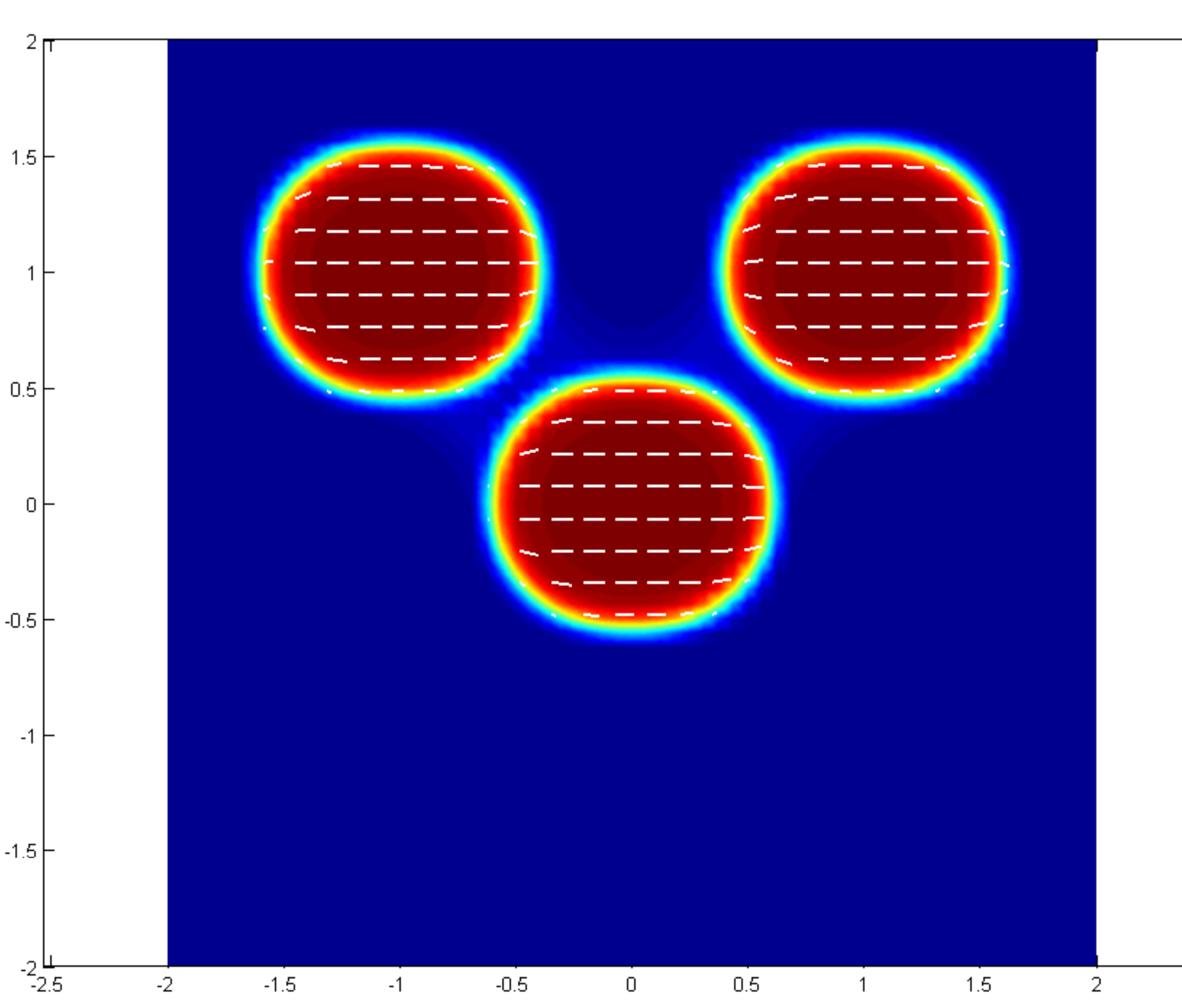}
\label{fig:results_phase_2}
} 
\subfigure[The tactoid shapes and director field at $t=0.2$. The tactoids begin to merge.]{
\includegraphics[width=0.45\textwidth]{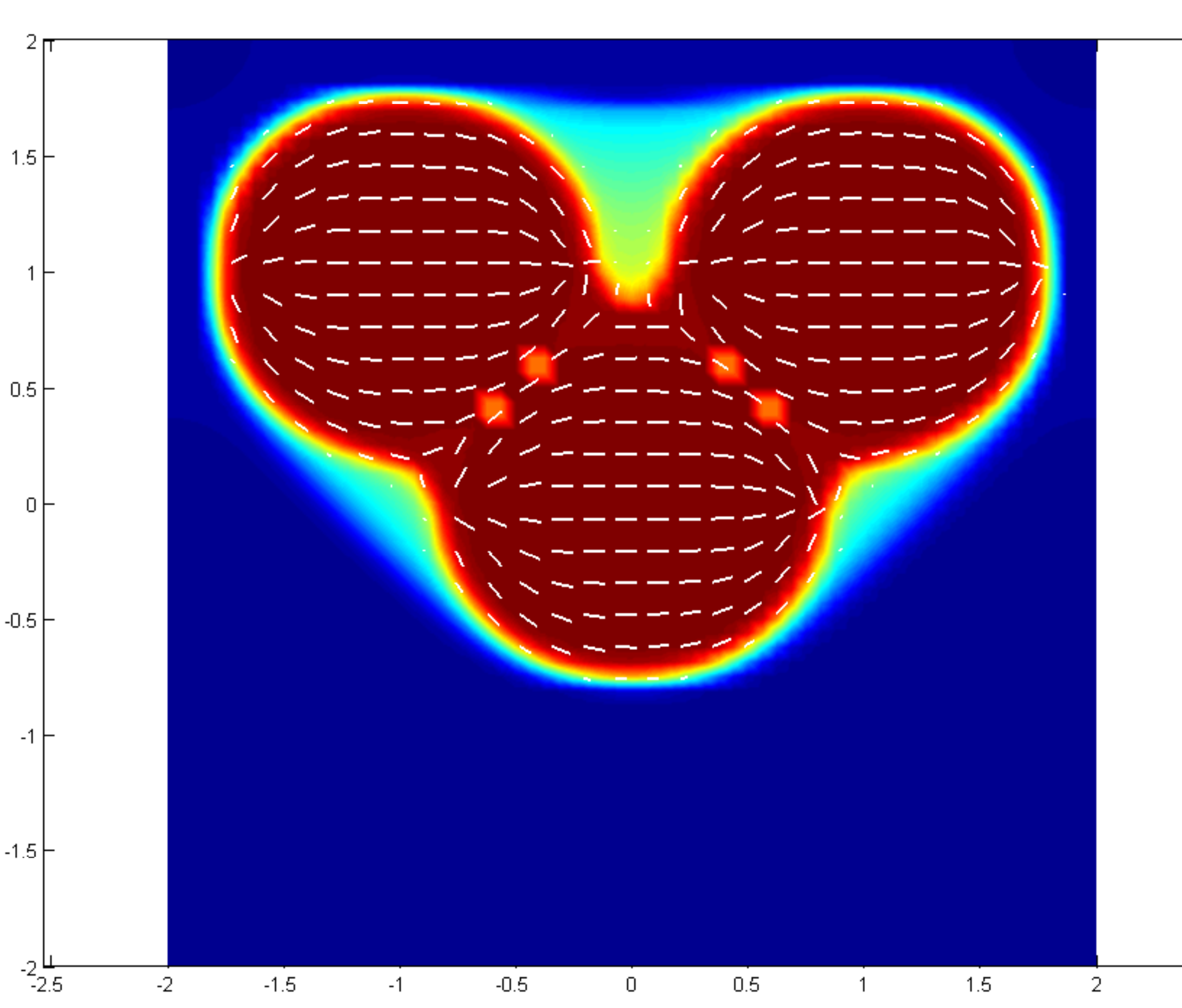}
\label{fig:results_phase_3}
}
\subfigure[The tactoid shapes and director field at $t=0.5$. a strength $-1$ disclination is formed inside the bulk.]{
\includegraphics[width=0.45\textwidth]{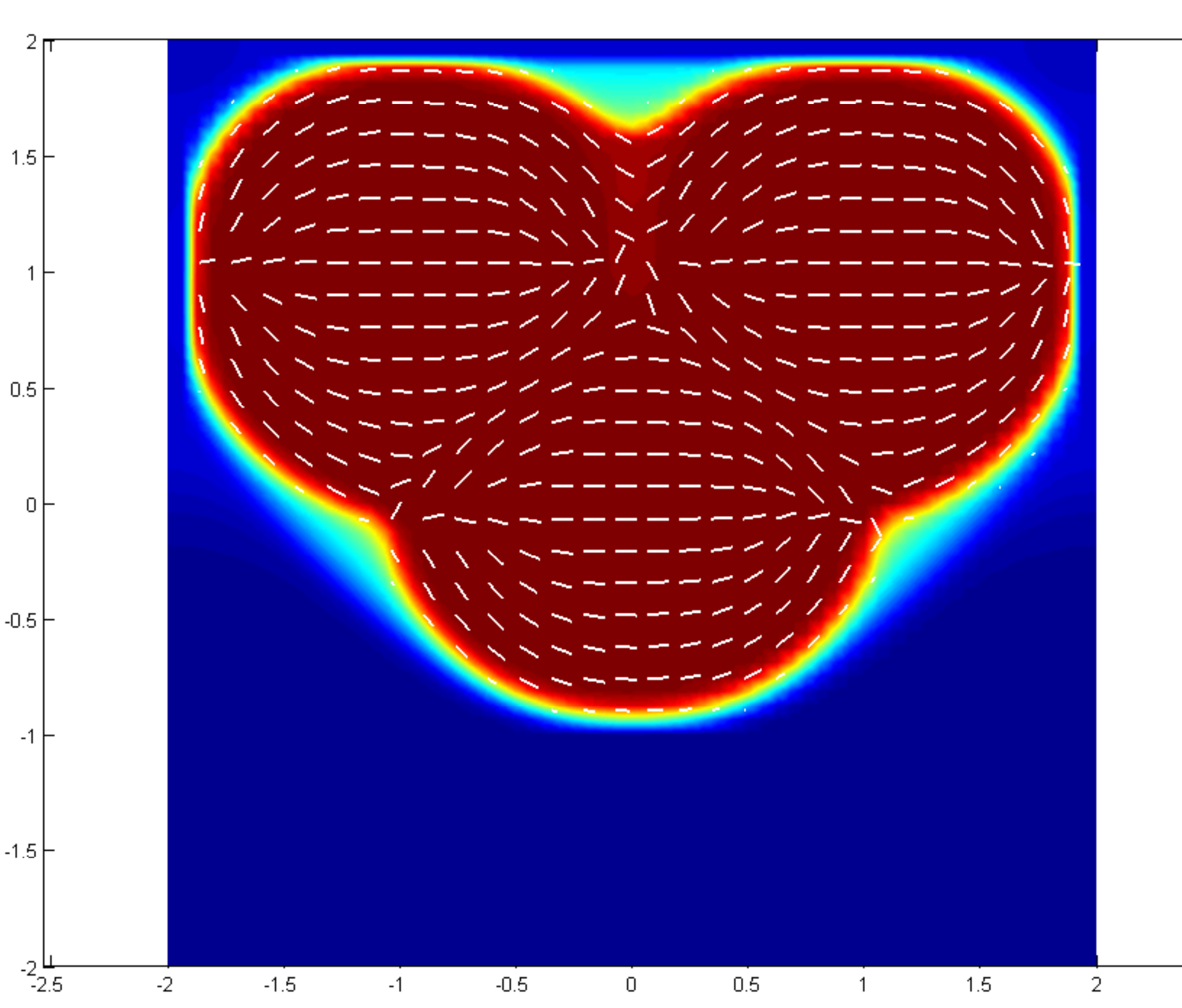}
\label{fig:results_phase_4}
}
\caption{Snapshots of isotropic-nematic phase transition at different times. As the calculation progresses, the tactoids expand, merge and a strength $-1$ disclination is formed inside the bulk.}
\end {figure}

We now discuss a problem of evolving phase transition across the whole domain. Three tactoids with different director orientations are initialized as shown in Fig. \ref{fig:results_phase_1}. The non-convex part $f(s)$ in the energy density is assumed to prefer the nematic phase, indicating the well at $s=1$ is lower than the well at $s=0$. The preference of the nematic phase of $f(s)$ indicates that the liquid crystal should transit from the isotropic to the nematic phase. Fig. \ref{fig:results_phase_2} to \ref{fig:results_phase_4} show snapshots at different times during the phase transition. As time increases, the tactoids expand and merge. In Fig. \ref{fig:results_phase_4}, a strength $-1$ disclination is formed inside the bulk which matches with experimental observations \cite{kim2013morphogenesis}. 

\section{Effect of material parameters on tactoid equilibria}\label{sec:para}

Since the energy proposed in this model is non-convex and the equilibrium of the tactoid and the director field depend on the interfacial energy and the Frank constants, it is of interest to explore tactoid equilibria as a function of material parameters.

\subsection{Frank constants $k_{11}$ and $k_{33}$}\label{sec:para_frank}

We consider two cases, $k_{11} > k_{33}$ (splay more expensive than bend) or $k_{11} < k_{33}$ (bend more expensive than splay). In one case, we assume $k_{11}$ is five time larger than $k_{33}$; in the other, we assume $k_{33}$ is five times larger than $k_{11}$. The tactoid shape is initialized as a sphere in both cases.

\begin{figure}
\centering
\subfigure[The initialized tactoid shape and director field.]{
\includegraphics[width=0.45\textwidth]{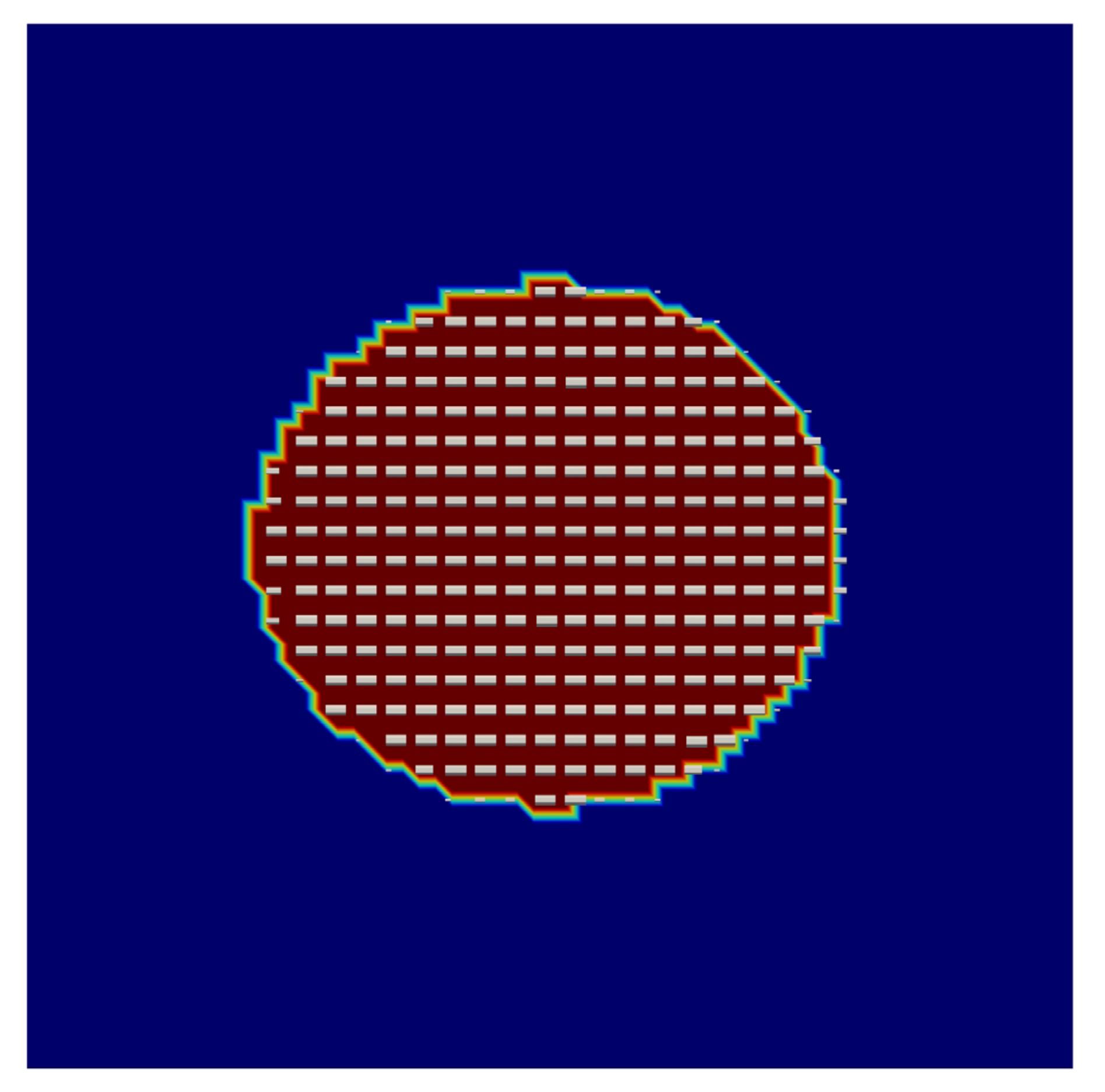}
\label{fig:frank_k11_1}
} 
\subfigure[The tactoid shape and director field at the equilibrium.]{
\includegraphics[width=0.45\textwidth]{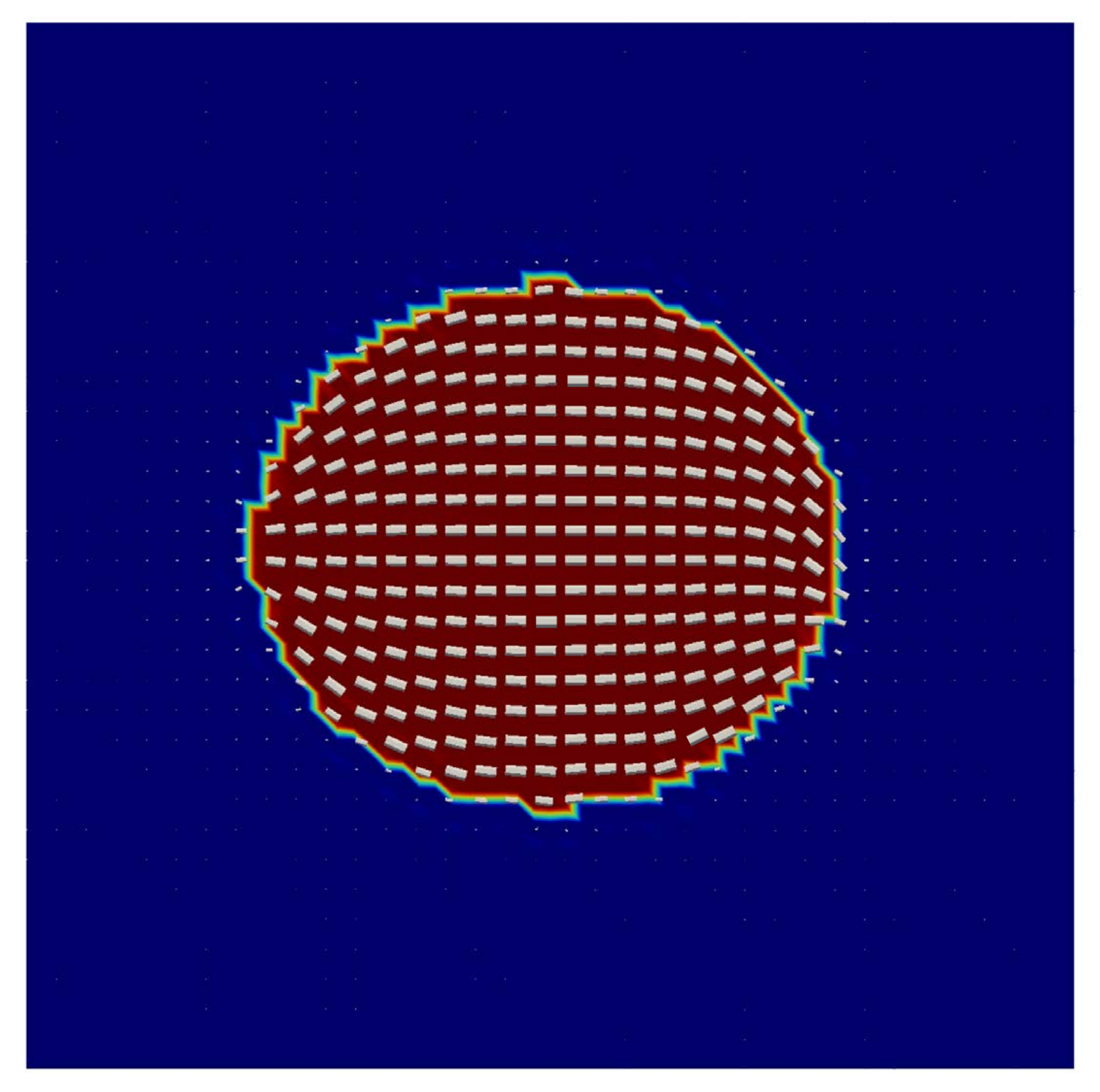}
\label{fig:frank_k11_2}
} 
\caption{The initialization and equilibrium configuration of the tactoid and director field in the case where $k_{11}>k_{33}$. Since splay is more expensive than bend, the director field tends to be perpendicular to the interface normal.}
\label{fig:frank_k11}
\end {figure}

\begin{figure}
\centering
\subfigure[The initialized tactoid shape and director field.]{
\includegraphics[width=0.45\textwidth]{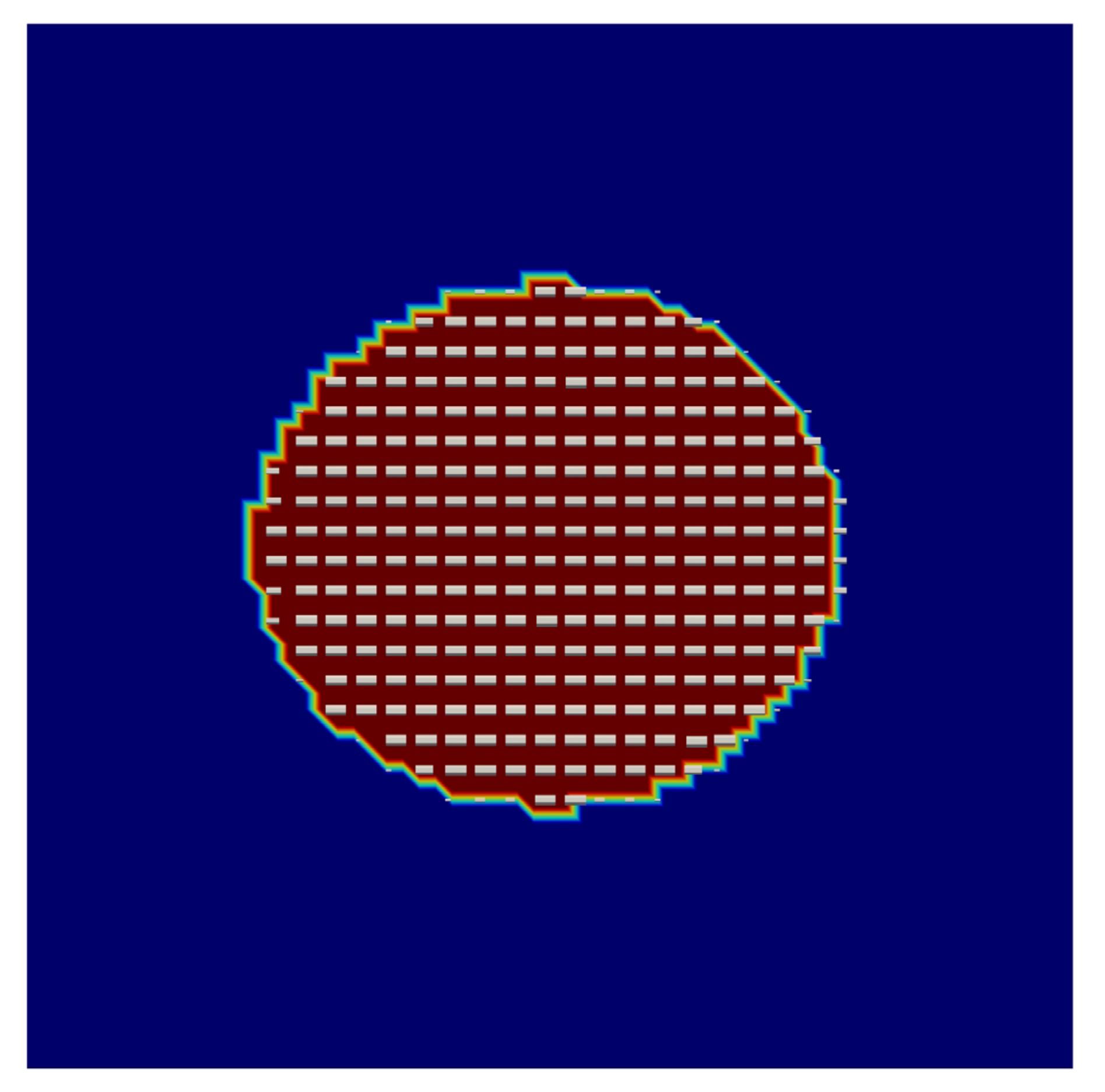}
\label{fig:frank_k33_1}
} 
\subfigure[The tactoid shape and director field at the equilibrium.]{
\includegraphics[width=0.45\textwidth]{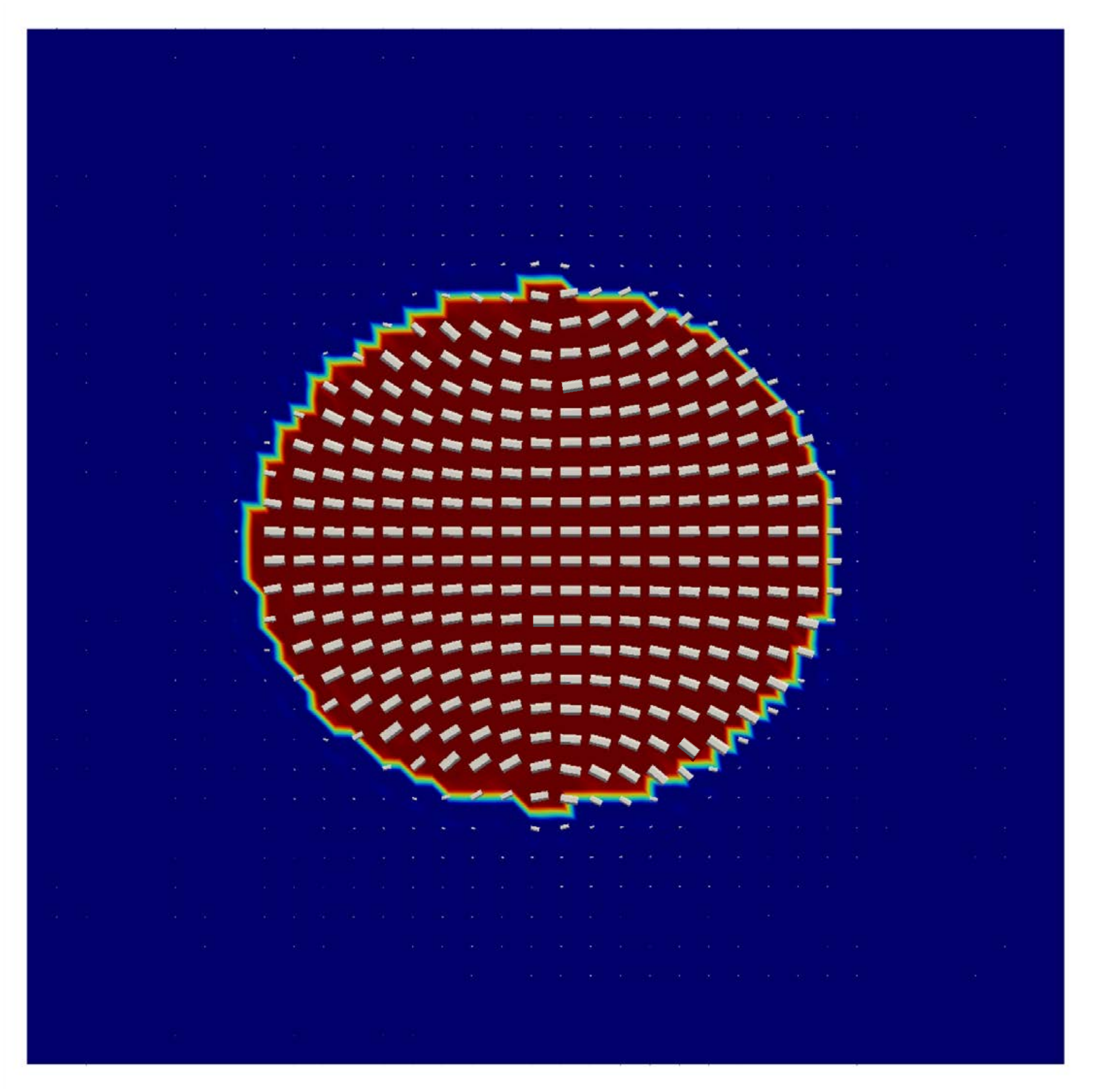}
\label{fig:frank_k33_2}
} 
\caption{The initialization and equilibrium configuration of the tactoid and director field in the case where $k_{11}<k_{33}$. Since bend is more expensive than splay, the director field tends to be parallel to the interface normal.}
\label{fig:frank_k33}
\end {figure}

Figs. \ref{fig:frank_k11} and \ref{fig:frank_k33} show the initial configuration and the equilibrium state for both cases. In Fig. \ref{fig:frank_k11}, $k_{11}$ is larger than $k_{33}$ and the director in the equilibrium tends to be perpendicular to the tactoid interface normal direction and bend is preferred over splay. On the other hand, in Fig. \ref{fig:frank_k33}, the director tends to be parallel to the interface normal direction with splay preferred over bend. The difference between these two results indicates that the relationship between $k_{11}$ and $k_{33}$ is crucial to the interaction between the director and the tactoid interface, which is also discussed in the  experiments reported in \cite{zhou2017ionic}.

\subsection{Effect of interfacial energy barrier on tactoid shape}

\begin{figure}
\centering
\includegraphics[width=0.6\textwidth]{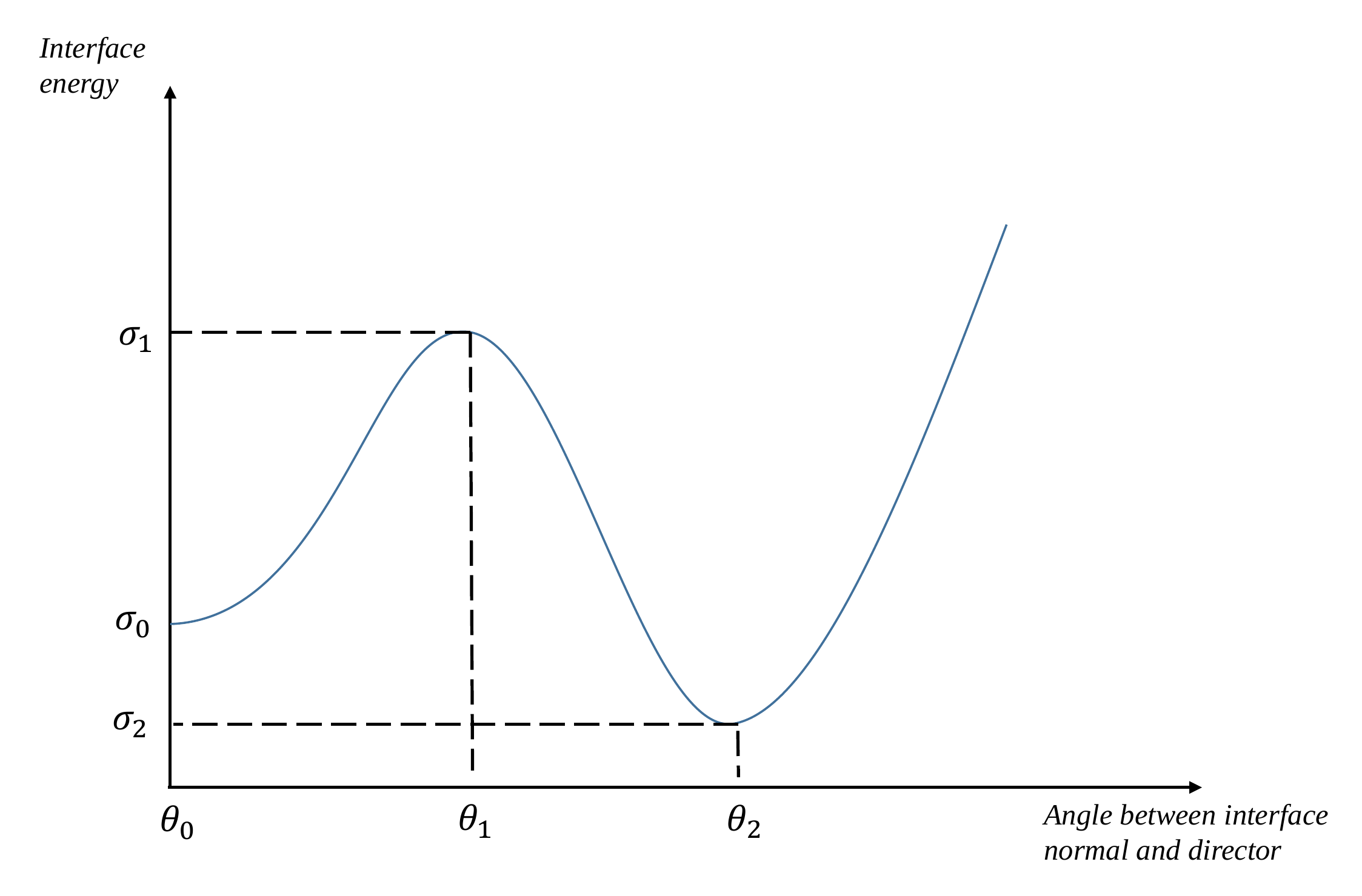}
\caption{The shape of the interfacial energy with two local minimal and a local maximal.}
\label{fig:interface_energy}
\end{figure}

Recall that in \eqref{eqn:interface_energy}, the interfacial energy is given in terms of the cosine of the angle $\theta$ between $\bfp$ representing the normal of the interface and the director field $\bfd$, which has a minimum at $\theta=\frac{\pi}{2}$. However, this approximation of the interfacial energy is only valid when the angle $\theta$ is close to $\frac{\pi}{2}$. We now assume an interfacial energy characterized by a fourth-order polynomial with two local minima and a local maximum as shown in Fig. \ref{fig:interface_energy}. Such a form of the surface anchoring potential was first introduced by Sluckin and Poniewierski \cite{Sluckin1986} and applied for the description of interfacial effects in LCLCs by Nazarenko et al \cite{nazarenko2010surface}. $\theta_0=0$ is where one local minimum occurs, $\theta_1$ is the location of the local maximum, and $\theta_2=\frac{\pi}{2}$ is the location of the other local minimum. $\sigma_0$, $\sigma_1$ and $\sigma_2$ are the interfacial energy values at $\theta_0$, $\theta_1$ and $\theta_2$, respectively. 

It is clear that the energy barrier between the two wells $\theta_0$ and $\theta_2$, as well as the values of $\sigma_0$ and $\sigma_1$, will influence the equilibrium state of the director field and the tactoid shape. Here we explore the relationship between the energy values of local maximum, as well as local minima, and the equilibrium of the director field. We assume $\theta_1=0.5$ and change $\sigma_0$, $\sigma_1$ and $\sigma_2$.

\begin{figure}
\centering
\includegraphics[height = 0.8\textheight]{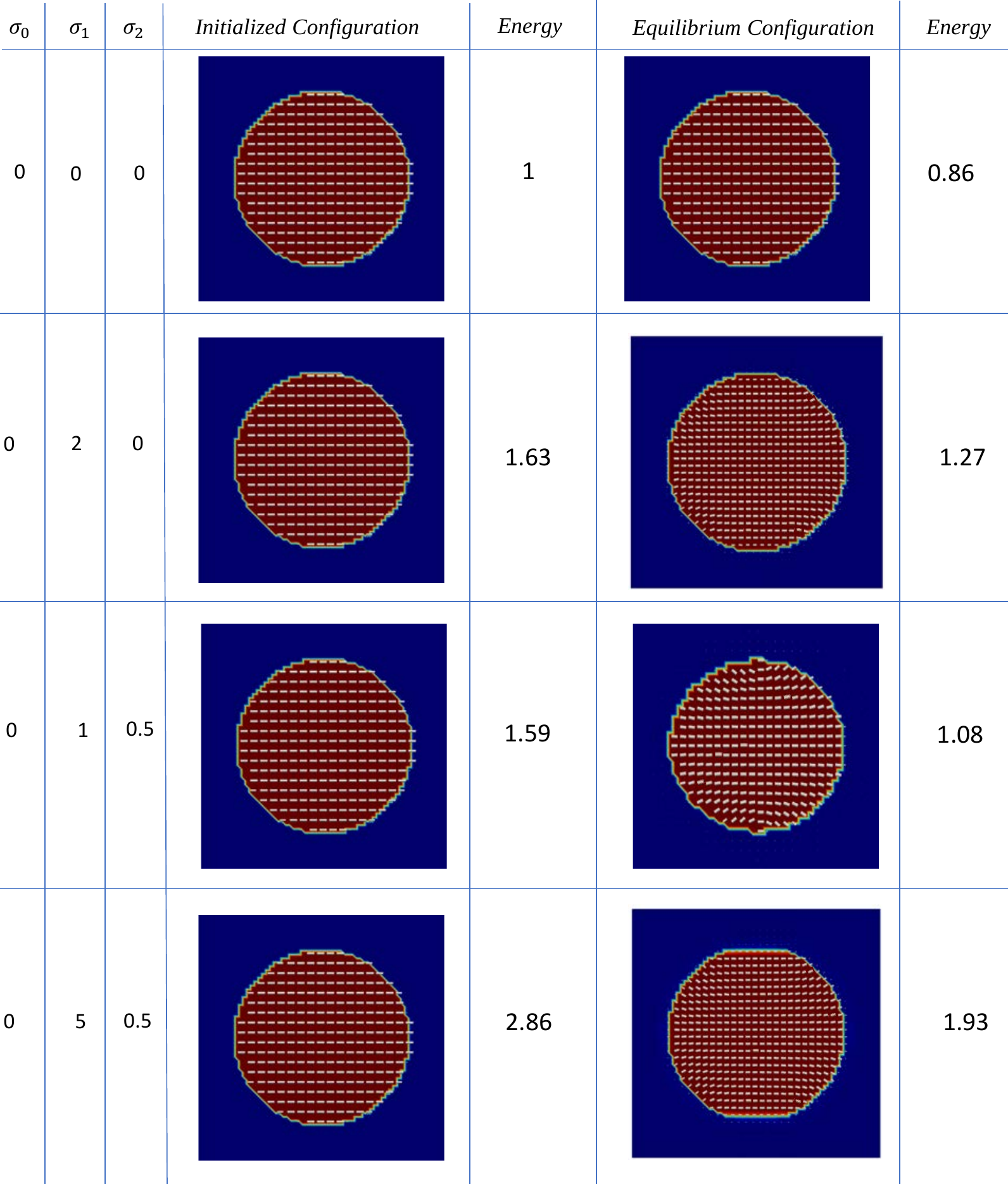}
\caption{The initializations and static equilibriums of the tactoid shape and director field given different interfacial energy parameters in the cases where $\sigma_0=\sigma_2$ and $\sigma_0<\sigma_2$. The total energy for each case are normalized by the energy value for the initialization of the case where $\sigma_0=\sigma_1=\sigma_2=0$.}
\label{fig:interface_table1}
\end{figure}

\begin{figure}
\centering
\includegraphics[height = 0.7\textheight]{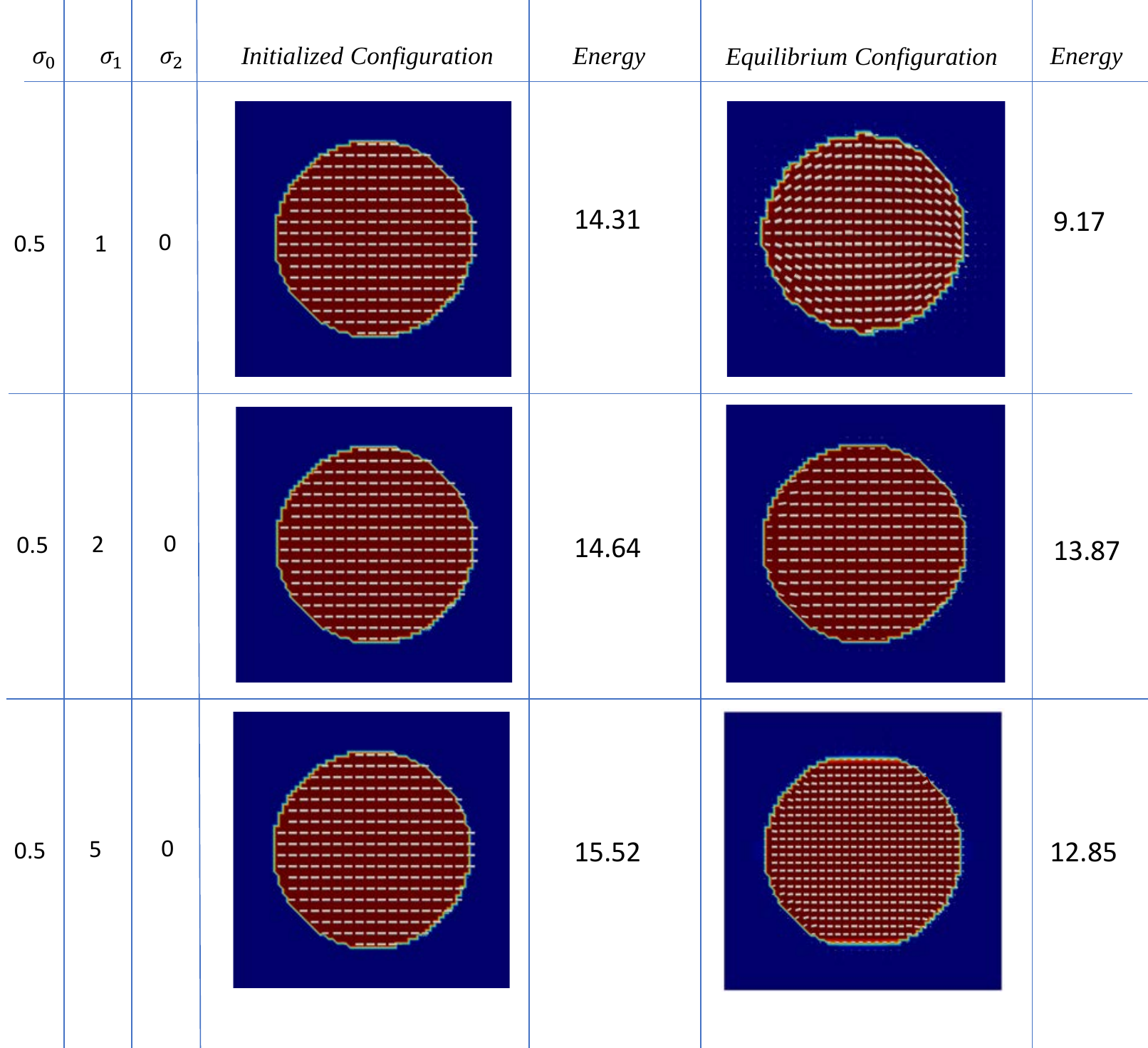}
\caption{The initializations and static equilibriums of the tactoid shape and director field given different interfacial energy parameters in the case where $\sigma_0>\sigma_2$. The total energy for each case are normalized by the energy value for the initialization of the case where $\sigma_0=\sigma_1=\sigma_2=0$.}
\label{fig:interface_table2}
\end{figure}

Fig. \ref{fig:interface_table1} and Fig. \ref{fig:interface_table2} show the initializations and equilibria of tactoid shapes and their director fields given different interfacial energy parameters. In the first row of Fig. \ref{fig:interface_table1}, $\sigma_0$, $\sigma_1$ and $\sigma_2$ are set to be $0$ so the interfacial energy will be zero at any angle between the director and the interface normal. Thus, the director field in the equilibrium is the same as the initialization. The second row shows the initialized configuration and static equilibrium corresponding to a higher $\sigma_1$ values. Since the energy barrier between $\theta_0$ and $\theta_2$ is high, the director field tends to move to its local minimum, namely some points being parallel to the interface normal and some points being perpendicular to the interface normal. The last two rows in Fig. \ref{fig:interface_table1} show different equilibria with the increasing energy barrier $\sigma_1$ in the case where $\sigma_0 < \sigma_2$. With low barrier $\sigma_1=1$, the director field can evolve to the lower well at $\theta_0$, thus the director field in the equilibrium is parallel to the interface normal. With high barrier $\sigma_1=5$, the director field cannot pass the local maximum between $\theta_0$ and $\theta_2$ and evolve to its local minimum in the equilibrium.  In addition, Fig. \ref{fig:interface_table1} shows the total energy for each case, which are normalized by the total energy of the case where $\sigma_0=\sigma_1=\sigma_2=0$.

Similarly, Fig. \ref{fig:interface_table2} shows the results with increasing energy barrier $\sigma_1$ in the case where $\sigma_0 > \sigma_2$. With low barrier $\sigma_1=1$, the director field can evolve to the lower well at $\theta_2$, and the director field in the equilibrium are perpendicular to the interface normal. With high barrier $\sigma_1=5$, the director field can only evolve to its local minimum in the equilibrium.  Fig. \ref{fig:interface_table2} also shows the total energy for each case, and the values of the total energy are normalized by the one of the case where $\sigma_0=\sigma_1=\sigma_2=0$.

\section{Conclusion} \label{sec:conclusion}

A model based in continuum kinematics and thermodynamics is derived for LCLC isotropic-nematic phase transition dynamics. By adopting the order parameter $s$ in \cite{ericksen1991liquid} to represent different phase states, an evolution equation of $s$ is proposed and discussed. The main difference between our model and Ericksen's model in \cite{ericksen1991liquid} is that the model in this work starts from a kinematic `tautology' with transparent physical/geometric motivation. The evolution of the director field described by the formulation in \cite{walkington2011numerical}. A new field $\bfp$ is introduced in the energy density to resolve the instabilities in the $s$ evolution resulting from  the non-convex interfacial energy when phrased only in terms of $grad s$ and $\bfd$. 

Both static equilibrium and dynamic tactoid behaviors are studied, including tactoid static microstructures from different initialized shapes, tactoid interactions, and isotropic-nematic phase transitions. The significance of the introduced evolution equation for $s$ is discussed in the context of describing tactoid dynamic behaviors. A parametric study is performed to explore the effect of nematic elastic constants (splay and bend) and the interfacial energy parameters on the interaction between the tactoid interface normal and the director field.

\section*{Acknlowledgment}
Support from the NSF DMREF program through grant DMS1434734 is gratefully acknowledged.

\newpage
\bibliography{bibtex}
\bibliographystyle{ieeetr}
\end{document}